\documentclass[10pt]{iopart}

\usepackage{graphicx}   
\begin{document}
\title{Bending of electric field lines and light-ray trajectories in a static gravitational field}
\author{Ashok K. Singal}
\address{Astronomy and Astrophysics Division, Physical Research Laboratory,
Navrangpura, Ahmedabad - 380 009, India }
\ead{ashokkumar.singal@gmail.com}
\vspace{10pt}
\begin{indented}
\item[]March 29, 2023
\end{indented}
\begin{abstract} 
It is well known that the light-ray trajectories follow a curved path in a gravitational field. This has been confirmed observationally where light rays coming from distant astronomical objects are seen to get bent in Sun's gravitational field. 
We explore here the bending of electric field lines due to gravity. We determine, from a theoretical perspective, not only the exact shapes of the bent trajectories of light rays, emitted isotropically by a source supported in a gravitational field, but also demonstrate that the electric field lines of a charge, supported in a gravitational field, follow exactly the trajectories of light rays emitted isotropically from a source at the charge location. 
From a detailed examination of the electrostatic field of a charge accelerated uniformly in the instantaneous rest frame, exploiting the strong principle of equivalence, we determine the bending of the electric field lines of a charge in a gravitational field.
The fraction of electric field lines crossing a surface, stationary above or below the charge in the gravitational field, are shown to be exactly similar to the fraction of light-ray trajectories intersecting that surface, emanating from a source lying at the charge location. 
On the other hand, for a freely falling charge in the gravitational field there is no such bending of electric field lines. The field lines continue to extend in radial straight lines from the instantaneous 'present' position of the charge, as do the trajectories of light rays spreading away from the  instantaneous position of a freely falling source in the gravitational field. 
The electric field configuration of a freely falling charge in the gravitational field is shown to be exactly the same as that of a charge moving uniformly in an inertial frame with velocity equal to the instantaneous ``present'' velocity of the freely falling charge. 
\end{abstract}
\ioptwocol
\section{Introduction}\label{SI}
A charge accelerated uniformly in its instantaneous inertial rest frame, and thus stationary in a comoving accelerated frame, can be considered, by the strong principle of equivalence \cite{EI20,MTW73,SC85,MO94,RI06}, as `supported' against gravity in a static gravitational field. Accordingly, from the electric field of a uniformly accelerated charge, which has a well-known solution in the literature \cite{5,10,JF15,88}, one can theoretically derive the electric field configuration of a charge stationary in a gravitational field. 
The electrostatic field distribution of such a charge shows systematic changes in the slopes of field lines. This bending of field lines can be attributed to the effect of gravity which leads to a finite electric field inside a uniformly charged spherical shell stationary in a gravitational field, contrary to what is expected otherwise. Further, we shall explicitly demonstrate that, due to gravitational bending of electric field lines of a stationary charge, the electric flux through a horizontal plane ``vertically above'' the charge in a gravitational field is less than that in a plane ``below'', which otherwise would have been equal in the absence of the gravitational field, according to Gauss law. 

It is well known from the general theory of relativity that a light ray, emitted along a straight horizontal path, bends `downward' in a gravitational field \cite{EI20,MO94,RI06}.
This has been amply confirmed observationally where light rays from distant astronomical objects coming along a tangential direction to Sun's limb, get bent in Sun's gravitational field, \cite{MTW73,RI06,Ho06,Dy19,9}. 
Here we shall determine the exact shapes of the bent trajectories of light rays, emitted  isotropically by a source, supported in a gravitational field. A comparison of the  electric field lines with the trajectories of light rays emitted from a source located at the charge position, will be made to demonstrate that the field lines exactly follow the trajectories of light rays in a static gravitational field. The total fraction of electric field lines crossing a surface, stationary in the gravitational field, would be calculated and shown thereby to be similar to the fraction of light-ray trajectories intersecting that surface. 
It will be further shown that at much `deeper' points in the gravitational field, the electric field lines, irrespective of their initial starting direction, turn almost vertical, similar to that for the trajectories of light rays. As the position of the stationary charge is chosen much deeper in the gravitational field, then an increasing amount of downward electric flux passes through a narrow region in the plane ``below'', and results in a much smaller amount of flux going upward through a similar region in the plane ``above''.  

One can also derive the electric field configuration of a freely falling charge in the gravitational field. The electric field in this case appears to be extending in radial straight lines from the instantaneous 'present' position of the charge, albeit bunched toward the horizontal plane perpendicular to the direction of free fall, exactly the same as the electric field of a charge moving uniformly in an inertial frame with a velocity equal to the instantaneous ``present'' velocity of the freely falling charge. From an explicit computation of Poynting vectors, it will be demonstrated that for a freely falling charge, no electromagnetic power gets radiated away from the charge. The net Poynting flow in this case turns out instead to be in the direction of movement of the charge, indicating a convective flow of energy in the fields surrounding the moving charge, essentially a 'downward' transport of the self-fields along with the freely falling charge. 
The electric field lines will appear exactly like the trajectories of light rays spreading away from the 'present' position of a freely falling source.
Actually the trajectories of individual light ray after being emitted from a freely falling source, as seen by  observers stationed in the comoving accelerated frame or in the gravitational field, may appear to be bent with respect to the time-retarded position of the emitting source. Nevertheless, the same observers will find the light rays to be spreading in radial straight lines with respect to the instantaneous position of the freely falling source, just like the electric field lines of a freely falling charge.
It will be further shown that for such a freely falling source there is no change in the total electric flux or in the total flux of light rays passing through a horizontal, infinite plane either above or below. 
\section{The equivalence of comoving frame of a uniformly accelerated charge to a frame stationary in static gravitational field}\label{S0}
We consider here a uniformly accelerated charge that progressively passes through a series of inertial frames as its successive instantaneous rest frames. We define thereby a comoving accelerated frame of the charge that coincides instantly with one of these inertial frames when the latter is momentarily the rest frame of the charge. Then the charge remains permanently stationary in the accelerated frame, which thus is its comoving frame. Then from the principle of equivalence we show how the comoving accelerated frame can be considered equal to a frame stationary in a static gravitational field and how the electromagnetic fields in the accelerated frame can be employed to compute electromagnetic fields of a charge stationary in the equivalent gravitational field.
\subsection{A comoving accelerated frame for a uniformly accelerated charge}\label{S0A}
In order to examine the electrostatic field of a charge held stationary in a static gravitational field, we begin with the case of a charge undergoing a uniform proper acceleration, say ${\bf a}_0$, along the $z$-axis. Such a charge progressively passes through a series of inertial frames as its successive instantaneous rest frames. The charge essentially is stationary in a comoving accelerated frame, say ${\cal A}$, with a series of inertial frames, successively, coinciding with ${\cal A}$. 

Let ${\cal I}, {\cal I}', {\cal I}'',\cdots$, be an infinite number of inertial frames, moving with relative velocities along the $z$-axis, the direction of acceleration, and one of them   coinciding with the comoving accelerated frame, ${\cal A}$, and thereby becoming the instantaneous rest frame of the uniformly accelerated charge. The spatial axes of various relatively moving inertial frames, ${\cal I}, {\cal I}', {\cal I}'',\cdots$, chosen to be parallel, coincide along with the origins $z=0,z'=0,z''=0,\cdots$, at times $t=0,t'=0,t''=0,\cdots$, and the standard Lorentz transformations relate the events in these reference frames \cite{MO94,11,83,RI06}.
Observers in the comoving accelerated frame will always find one of the infinite number of inertial frames, ${\cal I}, {\cal I}', {\cal I}'',\cdots$, to be instantaneously spatially coincident with their frame ${\cal A}$. Let ${\cal I}$ be one such instantaneous rest frame, which at time $t=0$ in ${\cal I}$ coincides with frame ${\cal A}$, also implying that the charge moving with a uniform acceleration  will instantly come to rest in the inertial frame ${\cal I}$ at $t=0$. Another frame ${\cal I}'$ will be instantaneously spatially coincident with ${\cal A}$ and thus be the instantaneous rest frame of the charge at $t'=0$ and so on \cite{11}.

It should be noted that events which are simultaneous in ${\cal I}$ are not simultaneous in ${\cal I}'$ and vice versa. In order for the successive inertial reference frames to be coincident with the comoving accelerated frame ${\cal A}$ of the charge, the value of acceleration would have to vary slightly along the $z$-axis. 
If $a_0$ is the magnitude of the acceleration in the comoving accelerated frame ${\cal A}$ at a reference point $z_{0}=c^{2}/a_{0}$, say, the location of the charge in ${\cal A}$, then the value of the acceleration at another point $z$ in ${\cal A}$ should be \cite{11,11a,83,17,RI06} 
\begin{eqnarray}
\label{eq:38.00}
a= \frac{c^2}{z}=a_0 \frac{z_0}{z}.
\end{eqnarray}
This spatial variation of acceleration (Eq.(\ref{eq:38.00})), known as Born rigidity condition, first derived by Max Born \cite{MB09}, ensures a simultaneous transition of all points ($z>0$) in the comoving accelerated frame from one instantaneous rest frame, say, ${\cal I}$, to the next instantaneous rest frame, say, ${\cal I}'$, preserving the  distances of separation between all points during the transition \cite{11,83,RI06,17}. 


By the equivalence principle, observers stationary in frame ${\cal A}$, comoving with the uniformly accelerated charge, can think of themselves ``supported'' in a static gravitational field, with the acceleration due to gravity, ${\bf g}=-{\bf a}$, the minus sign indicating that the acceleration due to gravity is along the negative $z$-axis.  Then from Eq.(\ref{eq:38.00}), the acceleration due to gravity $g$ at a point $z>0$ is related to the value $g_0$ at $z_0$, as
\begin{eqnarray}
\label{eq:38.01}
g=g_0 \frac{z_0}{z}.
\end{eqnarray}
In the equivalent gravitational field of  ${\cal A}$, the line element can be written as \cite{11a,17}
\begin{eqnarray}
\nonumber
{\rm d}s^{2}&=&-\left(\frac{g_0 z}{c}\right)^{2} {\rm d}t^{2}+{\rm d}z^{2}+{\rm d}\rho^{2}\\
\label{eq:38.02}
& =&-\left(\frac{z}{z_0}\right)^{2} c^2{\rm d}t^{2}+{\rm d}z^{2}+{\rm d}\rho^{2} 
\;.
\end{eqnarray}
A standard clock stationary at $z$, during a coordinate time interval ${\rm d}t$, measures an interval of proper time (``local'' time) ${\rm d}\tau=g_{0}z {\rm d}t/c^{2}$ \cite{MTW73,11}, whereas the coordinate time, by the convention adopted by us here, is the time measured on a standard clock at our chosen reference point, $z_{0}=c^{2}/g_{0}$ \cite{17}.

In order to understand how it works, let an inertial frame ${\cal I}$ be instantaneously spatially coincident with the comoving accelerated frame ${\cal A}$. Consider two points $z_1$ and $z_2$, fixed in ${\cal A}$, a distance $l$ apart along the $z$-axis, i.e., $z_2-z_1=l$, with $z_1$ lying deeper in the gravitational field. Then the ratio of the passage of the proper times at $z_1$ and $z_2$ in ${\cal A}$ is $(z_1/z_2)$, with the proper time passing slower at $z_1$. 

Let ${\cal I}'$ be another inertial frame moving relative to ${\cal I}$ with a small velocity $\Delta v$ along $z$-direction. Then observers in frame ${\cal I}$ will see instantaneously stationary points $z_1$ and $z_2$ gaining velocity with different accelerations, $a_1=c^2/z_1$ and $a_2=c^2/z_2$, so as to come to rest in frame ${\cal I}'$ after two different intervals of time, $\tau_1=\Delta v/a_1$ and $\tau_2=\Delta v/a_2$, with $\Delta\tau=\tau_2-\tau_1=\Delta v(z_2-z_1)/c^2$, as seen in  ${\cal I}$. However, $z_2$ and $z_1$, a distance $l$ apart in frame ${\cal I}$, will be seen by observers in frame ${\cal I}'$ to come to rest simultaneously with respect to their frame ${\cal I}'$, as from the Lorentz transformation  
\begin{eqnarray}
\label{eq:38.01a}
\Delta \tau'=\gamma(\Delta \tau-l \Delta v/c^2)=0\,.
\end{eqnarray}
This will be true for all pairs of points in  ${\cal A}$, implying all spatial points in  ${\cal A}$ would coincide with ${\cal I}'$ simultaneously. Thus observers in the comoving accelerated frame ${\cal A}$ will see that in a coordinate time interval $\Delta t=\Delta v/a_0$, or  in corresponding proper time intervals $\Delta \tau=\Delta v/a$ (Eq.~(\ref{eq:38.00})), the instantaneously spatially coincident inertial rest frame ${\cal I}$ will get replaced by another inertial rest frame ${\cal I}'$, moving relative to ${\cal I}$ with a  small velocity $\Delta v$ along $z$-direction. Such would keep happening with successive instantaneous inertial rest frame in turn coinciding with ${\cal A}$.
\subsection{Some peculiarities of the comoving accelerated frame and its relations with the set of  inertial frames including the instantaneous rest frame}\label{S0B}
Though for inertial frames ${\cal I}, {\cal I}', {\cal I}'',\cdots$, there is no size restriction and these inertial frames extend from $-\infty$ to $+\infty$ along the $z$ direction, however,  
the spatial extent of the comoving accelerated frame ${\cal A}$ is restricted to $z>0$ \cite{RI06}. The $z=0$ plane in the equivalent gravitational field is called an `event horizon', or simply horizon, as no signals from the $z<0$ regions can ever cross the $z=0$ plane to reach 
an observer stationary in the gravitational field, therefore such an observer cannot see beyond the horizon $z=0$. In fact as $z \rightarrow 0$, from Eq.(\ref{eq:38.01}), $g(z)\rightarrow \infty$ and the ratio of the proper time interval to the coordinate time interval, becomes zero. Any signal, therefore, from  the $z=0$ plane to reach an observer stationary at a finite $z$, say at $z_0$ in the gravitational field, would require an infinite time on the stationary observer's clocks. Moreover, in the comoving accelerated frame ${\cal A}$, the signals emitted in past, even at $t\rightarrow -\infty$ which will be from  $z\rightarrow \infty$, could at most be approaching the horizon ($z\rightarrow 0_+$), but it would never actually reach the horizon, $z=0$.

From the clock and length hypotheses \cite{RI06}, all momentarily space-time measurements by the accelerated observers will exactly match with those made in the instantaneous inertial rest frame. The instantaneous {\em proper time} interval measurements by the observers in the comoving accelerated frame ${\cal  A}$, will match those of clocks being carried by the observers in the instantaneously coincident inertial rest frame, say ${\cal I}'$. Also, while the length measured  in ${\cal A}$ of a meter rod lying in the instantaneous rest frame ${\cal I}'$, will be the same as in ${\cal I}'$, similar meter rods being carried in another inertial frames, say ${\cal I}$, having a Lorentz factor $\gamma$ of motion relative to the instantaneous rest frame ${\cal I}'$, will appear shorter in frame ${\cal A}$ by a factor $\gamma$ due to Lorentz contraction. Thus a point in ${\cal I}$ that earlier coincided with the charge position, $z_0$, in ${\cal  A}$, when ${\cal I}$ was the coincident frame for ${\cal  A}$, would now be at $z_0/\gamma$ in ${\cal  A}$. 
In fact, as seen in  ${\cal  A}$, the spatial dimensions of  ${\cal I}$ will continuously shrink, with the Lorentz factor $\gamma$ of the inertial frame  ${\cal I}$ increasing with respect to the changing instantaneous coincident frame of the comoving accelerated frame ${\cal  A}$. 

Thus, while any two given inertial frames will always be having a constant relative velocity, and hence the same Lorentz factor corresponding to their mutual velocity, the inertial frame coincident with the comoving accelerated frame ${\cal A}$ will be continuously changing. Therefore the velocity and Lorentz factor of various inertial frames relative to {\em that} inertial frame will also be continuously changing. Thus, observers in the comoving accelerated frame ${\cal A}$ will find the Lorentz contraction factor of various inertial frames  ${\cal I}, {\cal I}', {\cal I}'',\cdots$, to be continuously changing. Therefore, as seen by observers in  ${\cal A}$, the $z$-dimension of a frame, say ${\cal I}$, will first be linearly expanding as its Lorentz factor $\gamma$ relative to instantaneous rest frame of ${\cal A}$ would be becoming lesser till ${\cal I}$ itself becomes the instantaneous rest frame of ${\cal A}$ and thereafter the $z$-dimension of frame ${\cal I}$ will shrink as its Lorentz factor $\gamma$ relative to the next instantaneous rest frame of ${\cal A}$ would be becoming higher. In this scheme of things, the origins $z=z'=z''=0$ etc. of various inertial frames coincide with the origin point ($z=0$) of ${\cal A}$, in fact remain ever anchored to it as seen in  ${\cal A}$, since no temporal changes occur in ${\cal A}$ at the origin point, where the rate of passage of proper time is infinitely slow, and observers in  ${\cal A}$ can observe events only in $z>0$ regions, belonging to any of these inertial frames.

\subsection{Electromagnetic fields in the comoving accelerated frame versus in the equivalent gravitational field}\label{S0C}
 
From the strong principle of equivalence, the matching of measurements by the accelerated observers and those made by the observers in the equivalent gravitational field extend to those of the electromagnetic fields as well \cite{MTW73}. Thus we can calculate the electrostatic field of a charge held stationary in a static gravitational field, from the fields of a charge undergoing a uniform proper acceleration. An advantage of this also stems from that one can then apply results derived already in literature for the uniformly accelerated charge directly in the case of a charge stationary in a static gravitational field. In section~\ref{S1B} we show how the systematic changes in the slopes of electric field lines of such a charge, as seen in ${\cal A}$, allow us to compute the bending of field lines due to gravity \cite{18}. The bending of light-ray trajectories is computed in section~\ref{S2A}, and the effect of gravity on the bending of electric field lines and the light-ray trajectories is shown in section~\ref{S3} to be identical. The ``downward'' bending of electric field lines and light-ray trajectories, for supported sources in the gravitational field, makes the total electric flux as well as the  flux of lights ray passing through a horizontal plane ``above'' the source to be systematic less than the flux through an equivalent plane ``below''. In sections~\ref{S4A} and \ref{S4B},  this difference in the flux crossing the two planes is shown to be identical for electric field lines vis-\`e-vis light-ray trajectories, and in section~\ref{S5} we discuss the behaviour of electric field lines and the light-ray trajectories in the vicinity of the event horizon.

From an examination in ${\cal A}$ of the electromagnetic fields of a charge stationary in an inertial frame, we find no bending in the fields of a charge freely falling in the gravitational field (section~\ref{S1C}). The electric field lines of such a charge in frame ${\cal A}$ are radial everywhere, albeit concentrated towards the plane perpendicular to the direction of motion. The latter is consistent with all $z$-dimensions of frame ${\cal I}$ Lorentz contracted with respect to the instantaneous rest frame of ${\cal A}$. Similar, of course, is the behaviour of light-ray trajectories from a source freely falling in the gravitational field (section~\ref{S2B}).

One can understand the behaviour of field lines, directly, from the perspective of observers stationary in the gravitational field. From the principle of equivalence, {\em all things}, including the matter as well as the associated fields, fall in a gravitational field. Therefore, as a charged particle undergoes a free fall, so does the bundle of 
electric field lines along side it, for {\em field points at all distances} in the horizontal plane containing the charge \cite{17}. 
Seen this way, it can be easily understood, why no electromagnetic radiation, whatsoever, from a freely falling charge in a uniform gravitational field would take place, contrary to the expectation in the standard formulation of radiation from a charge accelerated ordinarily with respect to an inertial frame \cite{1,2,25,PU85}.

On the other hand a charge having a constant proper acceleration too emits no radiation, because it turns out (section~\ref{S1A}) that the magnetic field is zero {\em everywhere} in its instantaneous rest frame. Moreover, by the strong principle of equivalence it is equivalent to a charge permanently stationary in a gravitational field \cite{4,RI06}, which is a time-static situation. This condition of no temporal variations would otherwise be violated, both in the charge as well as in the electromagnetic fields around it, if in this case there were radiation taking place from the charge. However, at far-off distances from the time-retarded positions of a uniformly accelerated charge, when the charge is moving with a finite velocity, one does find a finite Poynting flux, which has been taken as evidence of radiation \cite{5,10,PA02,AL06}.
However, it has been later shown explicitly that when the leading spherical front of the relativistically beamed Poynting flux, advances forward at a large time to a far-off distance, the uniformly accelerated charge too is not lagging far behind. In fact, these relativistically beamed fields, increasingly resemble fields of a charge moving in an inertial frame with a uniform velocity equal to the instantaneous 'present' velocity of the uniformly accelerated charge, 
with a convective flow of fields in that frame along with the movement of the charge \cite{89}. There is no other Poynting flow in the far-zones that could be termed as {\em radiation emitted} by a uniformly accelerated charge and this, in turn, is fully consistent with not only the absence of radiation reaction on such a charge but is also fully conversant with the strong principle of equivalence. 

\section{Electric field of a charge in a static gravitational field}\label{S1}
\subsection{A charge `supported' in a 'uniform' gravitational field}\label{S1A}
Let a charge $e$, undergoing a uniform acceleration $a_0$ along the $z$-axis comes to rest momentarily at $z_0=c^{2}/a_0$ at time $t=0$ in the inertial frame ${\cal I}$.  
The electric field, expressed in cylindrical coordinates $(\rho,\phi,z)$, at a point in the $(\rho,z)$ plane ($\phi=0$) in the instantaneous rest frame of the charge is given by \cite{5,10,JF15,88} 
\begin{eqnarray}
\label{eq:38.ab1}
E_{\rho}&=&\frac{8ez_0^{2}\rho z}{\xi^{3}}\nonumber\\
E_{\rm z}&=& \frac{4ez_0^{2}(z^{2}-z_0^{2}-\rho^{2})}{\xi^{3}}\,,
\end{eqnarray}
where 
$\xi={[(z_0^{2}-z^{2}-\rho^{2})^2+4z_0^{2}\rho^{2}]^{1/2}}$.
The remaining  field components, including the magnetic field, are everywhere zero. 
From causality condition, the field  (Eq.~(\ref{eq:38.ab1})) exists only in the region $z>0$ \cite{5}. Since there is no magnetic field, the Poynting flux is nil everywhere, accordingly there is no radiation from a uniformly accelerated charge, as first pointed out by Pauli \cite{33}.

The expression for electromagnetic fields (Eq.~(\ref{eq:38.ab1})), can be written 
in polar coordinates ($R,\psi,\phi$), centered on the charge position at $z_0$ \cite{18,89,24}, as 
\begin{eqnarray}
\label{eq:38.1a}
z=z_{0}+R \cos\psi, \,\,\,\,\,\, \rho=R \sin\psi\,,
\end{eqnarray}
where $\psi$ is the angle with respect to the direction of acceleration, assumed to be along the $z$-axis, and we assume $\phi=0$. 
Substituting in Eq.~(\ref{eq:38.ab1}), and after some algebraic simplifications, we get
\begin{eqnarray}
\nonumber
E_{\rm R}=\frac{e(1+\eta \cos\psi)}{R^{2}(1+2\eta\cos\psi +\eta^{2})^{3/2}}\\
\label{eq:38.ab2}
E_{\psi}=\frac{e\eta \sin\psi}{R^{2}(1+2\eta\cos\psi +\eta^{2})^{3/2}}\,,
\end{eqnarray}
with $\eta=a_0R/2 c^{2}=R/2z_0$. All the remaining field components are zero. In the absence of acceleration ($a_0=0$), $\eta=0$ and from Eq.~(\ref{eq:38.ab2}), the field reduces to radial Coulomb field of a charge, $E_{\rm R}=e/R^2$, with other field components zero. 

Equation~(\ref{eq:38.ab2}), or alternatively Eq.~(\ref{eq:38.ab1}), gives the electric field in the instantaneous rest frame ${\cal I}$ of the uniformly accelerated charge, which By clock and length hypotheses \cite{RI06}, 
is also the electric field in the comoving accelerated frame ${\cal A}$. Then from the strong principle of equivalence, Eq.~(\ref{eq:38.ab2}) (or alternatively Eq.~(\ref{eq:38.ab1})) gives the electrostatic field of a charge ``supported'' at $z_0$ in a static gravitational field, with the acceleration due to gravity $g_0=a_0$ at $z_0$, while at a general point $z$ the acceleration due to gravity is given by Eq.~(\ref{eq:38.01}). Of course, everywhere the direction of the gravity is along the negative $z$-axis.

A 2-d analysis of the electric field lines in the $\rho-z$ plane is provided in section~\ref{S1B}, especially in Figs.~\ref{F1} and \ref{F2}.
\subsection{Bending of the electrostatic field lines of a charge {\em stationary} in a gravitational field}\label{S1B}
\begin{figure}[t]
\begin{center}
\includegraphics[width=\columnwidth]{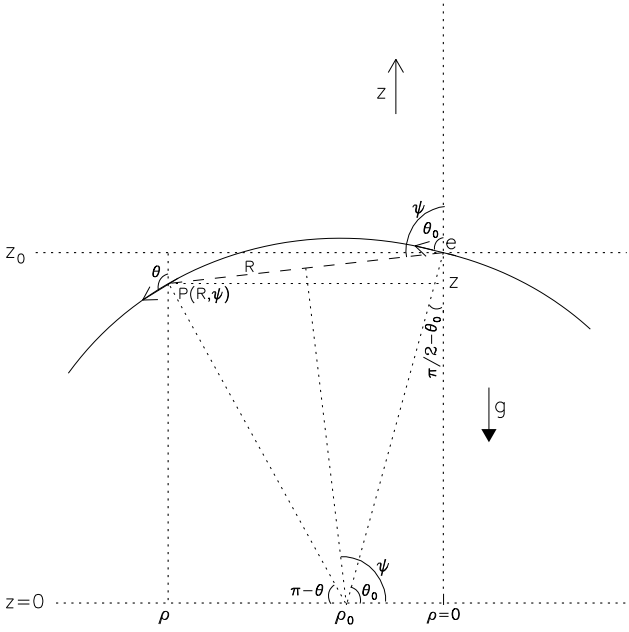}
\caption{Electric field lines emanating from the charge $e$, stationary in the gravity field at $\rho =0,z=z_0$, bend due to the gravitational field in the shape of circles, centered at $\rho=\rho_{0}$ on the $z=0$ plane with radii $(\rho_0^2+z_0^2)^{1/2}$. A typical electric field line starting along $\theta_0$ from the charge at $z_0$ is along $\theta$ at a field point $(\rho,z)$ 
(or $P(R,\psi)$ in polar coordinates with origin on the charge position at $z_0$).}
\label{F1}
\end{center}
\end{figure}
\begin{figure}[t]
\begin{center}
\includegraphics[width=\columnwidth]{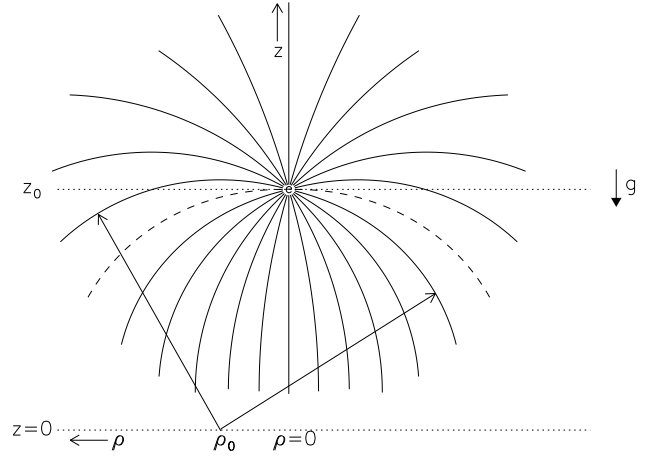}
\caption{Gravitational bending of the electric field lines of a charge $e$, held stationary at  say, $\rho=0,z=z_0$, in the gravitational field. A typical electric field line, starting radially from the charge, is a circle of radius $\sqrt{\rho_{0}^{2}+z_0^{2}}$, centered at $\rho=\rho_0,z=0$, thereby passing through the charge position.}
\label{F2}
\end{center}
\end{figure}
The electrostatic field lines of a charge, permanently stationary in the gravitational field, start in radial directions from the charge position, progressively bend 'downward' due to gravity when seen at increasing radial distances from the charge. The bending of electric field lines leads to some curious outcome, unexpected otherwise, like 
the presence of a finite electric field inside a uniformly charged spherical shell stationary in a gravitational field, a suitable experimental setup of which, with a clever design, might make it a possible test of the strong principle of equivalence (see \ref{SAA}). 

From Eq.~(\ref{eq:38.ab2}), for small $R$, when $\eta\ll 1$, we have
\begin{eqnarray}
\label{eq:38.ab3}
E_{\rm R}=\frac{e}{R^{2}}(1-2\eta \cos\psi)\,,\,\,\,\,\,
E_{\psi}=\frac{e \eta \sin\psi}{R^2}\,.
\end{eqnarray}
Thus a stationary charge, which has only a radial electric field in an inertial frame, when held in a gravitational field, possesses to a first order, a finite non-radial electric field 
\begin{eqnarray}
\label{eq:38.abc3}
\frac{E_\psi}{E_{\rm R}}\sim \eta \sin\psi=\frac{R \sin\psi}{2z_0}= \frac{g_0 R \sin\psi}{2c^2}\,,
\end{eqnarray}
even in the vicinity of the charge, which shows the bending of the electrostatic field lines due to the gravitational field.

In order to explore more exactly the electrostatic field configuration of a charge in a gravitational field, we recall that the electric field vector at any spatial location represents a tangent 
to the electric field line through that point. 
Then, from Eq.~(\ref{eq:38.ab2}), we can write the differential equation for a field line through a field point $(R,\psi)$ as 
\begin{eqnarray}
\label{eq:38.2a}
\frac{1}{R}\frac{{\rm d}R}{{\rm d}\psi} 
=\frac{E_{\rm R}}{E_{\psi}} =\frac{2z_0+R \cos\psi}{R\sin\psi}\,,
\end{eqnarray}
Equation~(\ref{eq:38.2a}) has a solution 
\begin{eqnarray}
\label{eq:38.2a1}
z_0\cot\psi + (R/2)\csc\psi=\rho_0\,,
\end{eqnarray}
where $\rho_0=z_0\cot\theta_0$ is a constant of integration, which specifies for the electric field line passing through the field point ($R,\psi$), its initial radial direction, emanating from the charge position at $z_0$, since for $R\rightarrow0$, from Eq.~(\ref{eq:38.2a1}), $\psi\rightarrow \theta_0$. 

Equation~(\ref{eq:38.2a1}) can be rewritten as
\begin{eqnarray}
\label{eq:38.2a2}
R=2(z_0\csc\theta_0)\sin(\psi-\theta_0)\,,
\end{eqnarray}
which is the polar equation of a circle of radius $z_0 \csc\theta_0$, that passes through the pole at $z_0$, with the radius from the center of the circle to the pole $z_0$ making an angle $\pi/2-\theta_0$ with the $z$-axis (Fig.~\ref{F1}). 

By writing the radius $z_0\csc\theta_0=\sqrt{\rho_{0}^{2}+z_0^{2}}$, we can rewrite the Eq.~(\ref{eq:38.2a1}) for the field line in the following, a more readily recognized, form 
\begin{eqnarray}
\label{eq:38.2a3}
(R\sin \psi-\rho_{0})^{2}+(z_0+R \cos\psi)^{2}=\rho_{0}^{2}+z_0^{2}\,,
\end{eqnarray}
or using Eq.~(\ref{eq:38.1a})
\begin{eqnarray}
\label{eq:38.2a4}
(\rho-\rho_{0})^{2}+z^{2}=\rho_{0}^{2}+z_0^{2}\,,
\end{eqnarray}
which is a circle with center at $(\rho=\rho_{0},z=0)$ and of a radius 
$(\rho_0^2+z_0^2)^{1/2}$, and thus passing through the charge at ($\rho =0,z=z_0)$. Geometrical relations between various quantities in Eqs.~(\ref{eq:38.2a2}), (\ref{eq:38.2a3}) or (\ref{eq:38.2a4}) can be visualized from Fig.~\ref{F1}.

We can plot the field lines by choosing a point $\rho_{0}$, arbitrarily on the $z=0$ plane as a center and draw a circle which passes through the charge. Different field lines, starting along different initial radial directions ($\theta_0$) from the charge position at $z_0$, are obtained by varying the position of $\rho_{0}$ on the $z=0$ plane. 

Figure~\ref{F2} shows a plot of the field lines.
It appears that the field lines start radially away from the charge position 
but are continuously `bent' by the gravitational field, becoming vertical as one approaches the $z=0$ plane.
This perhaps is the simplest example of the bending of the electrostatic field lines
of a charge by a gravitational field. 
\subsection{A charge {\em freely falling} in a 'uniform' gravitational field}\label{S1C}
\begin{figure}[t]
\begin{center}
\includegraphics[width=\columnwidth]{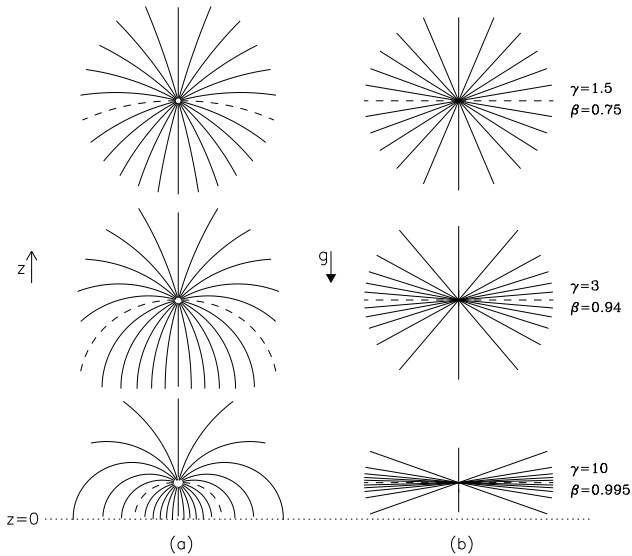}
\caption{A schematic representations of the electric field lines of (a) charges stationary with respect to the local observers, supported at different depths in a `uniform' gravitational field (b) a freely falling charge at different instants, as it gathers speed with respect to the observers stationary in the gravitational field. The dashed lines represent the field line starting from the charge in the horizontal plane, i.e. in a plane normal to the $z$-direction.}
\label{F4}
\end{center}
\end{figure}

A schematic of the electrostatic field lines of a charges stationary with respect to the local observers, supported at different depths in a `uniform' gravitational field, is presented in Fig.~\ref{F4}a, which shows that the bending of the electrostatic field lines becomes more pronounced as the static location of charge is chosen closer to the horizon at $z=0$.

On the other hand, for a charge freely falling along the $-z$ direction in the uniform gravitational field, there is none whatsoever bending of the electric field lines. The electric field configuration for a freely falling charge is shown in Fig.~\ref{F4}b (also see \cite{17,Er04}), where due to the free fall of the charge as well as of
its associated fields, the electric field lines are always along radial straight lines
from the instantaneous position of the charge, though they increasing bunch perpendicular to the direction of motion as the charge gathers speed during its free fall due to gravity. 

A charge freely falling in a uniform gravitational field is actually stationary in a co-falling  inertial frame, say ${\cal I}$, where it has a Coulomb field and as the charge  freely falls in the gravitational field, so does its inertial rest frame ${\cal I}$, along with the bundle of Coulomb field lines around the charge. Thus the electric field lines remain radial from the instantaneous position of the charge freely falling in a uniform gravitational field. However, as the charge, due to the acceleration of gravity along the $-z$ direction, gathers speed along with ${\cal I}$, another inertial frame, say ${\cal I}'$, becomes the instantaneous coincident frame with ${\cal A}$,  the electric field lines at that instant in ${\cal A}$, will appear to be the same as seen in ${\cal I}'$, concentrated towards the plane perpendicular to the direction of motion \cite{1,PU85,25}, as if radial lines, emanating from the ``present'' position of the charge, were contracted in the direction of motion by the instantaneous Lorentz factor of the velocity arising from its free fall. 
Effectively, to the observer stationary in the gravitational field, there is a Lorentz contraction of distances between horizontal planes of  ${\cal I}$ (section \ref{S0B}). Therefore, as seen by the observer stationary in the gravitational field, there is a bunching of the field lines toward the horizontal plane passing through the instantaneous position of the freely falling charge. 

Since the system has a cylindrical symmetry, it is convenient to employ cylindrical coordinates ($\rho,\phi,z$) in the case of a freely falling charge to describe the electromagnetic field components, which  in the $(\rho,z)$ plane are given by \cite{17}
\begin{eqnarray}
\label{eq:38.03}
E_{\rho}  =  \frac{e \gamma \rho}{[\gamma^{2}(\Delta z)^{2}+\rho^{2}]^{3/2}} 
\nonumber \\
E_{z}  =  \frac{e \gamma \Delta z}{[\gamma^{2}(\Delta z)^{2}+\rho^{2}]^{3/2}} 
\nonumber \\
B_{\phi}  =  \frac{-e \gamma \rho v/c}{[\gamma^{2}(\Delta z)^{2}+\rho^{2}]^{3/2}} \;,
\end{eqnarray}
with all other field components being zero. Here $\Delta z$ represents the
distance of the field point along the $z$-axis with respect to the ``present'' position, $z_{\rm e}$, of the freely falling charge, $v$ is the magnitude of the``present'' velocity of the charge during its free fall and and $\gamma= 1/\sqrt{1-(v/c)^2}$ is the corresponding Lorentz factor. 

We can rewrite the electromagnetic fields in Eq.~(\ref{eq:38.03}) as
\begin{eqnarray}
\label{eq:38.03a}
{\bf E}&=&\frac{e{\bf R}}{\gamma^2[(\Delta z)^{2}+\{1-(v/c)^{2}\}\rho^2]^{3/2}}\nonumber\\
&=&\frac{e\hat{\bf R}}{R^{2}\gamma^{2}[1-(v/c)^{2}\sin^{2}\psi]^{3/2}}\,,
\end{eqnarray}
with the magnetic field, ${\bf B}= {\bf v}\times {\bf E}/c$, where ($R,\psi,\phi$) are polar coordinates of the field point with respect to the charge position at $z_{\rm e}$. The instantaneous field for the freely falling charge, given by Eq.~(\ref{eq:38.03a}), is exactly the same as the field of a charge moving with a uniform velocity $v$ \cite{1,2,25,PU85}.

The electric field configuration for a supported charge (Fig.~\ref{F4}a, also see \cite{18}) thus is quite different from that of a freely falling charge (Fig.~\ref{F4}b).
The electric field lines of a charge, supported in a gravitational field, start radially from the charge position, however, unlike for a freely falling charge, continuously bend in the ``downward'' direction, due to the effect of gravity (Fig.~\ref{F4}a). 

On the other hand, the electric field lines of a freely falling charge are exactly similar to those of a charge moving with a uniform velocity $v$ equal to the instantaneous velocity of free fall of the charge and its corresponding Lorentz factor $\gamma$. The field lines are centered on the ``present'' position of the free falling charge, and as its $\gamma$ becomes larger due to the increasing velocity of free fall, the electric field component along the direction of motion, becomes negligible relative to the perpendicular component, with field lines increasingly getting oriented perpendicular to the direction of motion. Moreover, the field around the charge in regions along the direction of motion, is appreciable only in a narrow zone that shrinks as $\Delta z\propto 1/\gamma$, for large $\gamma$ \cite{88,PU85}. 

A question could arise here. For a charge moving with a uniform velocity there is no radiated power being emitted. What about a freely falling charge, which is constantly accelerated due to gravity with respect to an observer stationary in the gravitational field? This is discussed in \ref{SAB}, where it is shown that though there exists a finite Poynting flux, nonetheless there is no radiation going out from the charge in either case.  
\section{Trajectories of light rays originating from a source in the gravitational field}\label{S2}
\subsection{For a source {\em stationary} in the gravitational field}\label{S2A}
A horizontal beam of light rays from a source stationary in an inertial frame, would appear to be bending `downward' as seen from a frame accelerated `upward'. From this, using the equivalence principle, it has been inferred that the horizontal trajectory of a light ray gets bent due to gravity \cite{EI20,MO94,RI06}.
We want to determine the exact shapes of the bent trajectories of light rays, emitted initially isotropically by a source $\cal S$, supported in a gravitational field. For this we exploit the equivalence principle and compute the trajectories of light rays emitted from $\cal S$,  assumed to be stationary in the comoving accelerated frame ${\cal A}$, and accordingly accelerated uniformly with respect to an inertial frame. 

Consider a light ray emitted along direction $\theta_0$, measured with respect to the $z$-axis, from the source ${\cal S}$, assumed to be stationary at position $z_0=(c^2/a)$ in the comoving accelerating frame ${\cal A}$. We designate that event by ${\cal E}_1$. Let ${\cal I}'$ be an inertial frame, momentarily coincident with the frame ${\cal A}$, where at time $t'=0$ of the inertial frame ${\cal I}'$ this light ray is emitted along direction $\theta'$. Since ${\cal I}'$ at time $t'=0$ is the instantaneous rest frame, coincident with the comoving accelerating frame ${\cal A}$, then the light ray emitted along $\theta'$ in ${\cal I}'$ would be along $\theta_0$ in frame ${\cal A}$ with $\theta_0=\theta'$.

Light will continue to move in a straight line along $\theta'$ in the inertial frame ${\cal I}'$. Let ${\cal E}_2$ be the event where light, after a time duration $\Delta t'$, arrives at a point $c\Delta t'$ with 
\begin{eqnarray}
 \nonumber
 \Delta z'&=&c\Delta t' \cos\theta'\\
\label{eq:38.ab5a}
\Delta \rho'&=&c\Delta t' \sin\theta'\,.
\end{eqnarray}

The frame ${\cal I}'$ will no longer be coinciding with ${\cal A}$ at event ${\cal E}_2$. Instead, let another inertial frame ${\cal I}$ be the one that coincides with the comoving accelerating frame ${\cal A}$ and thus will be the instantaneous rest frame of the source ${\cal S}$, when event ${\cal E}_2$ takes place, i.e. $t_2=0$. Let frame ${\cal I}$ be moving with a velocity $v$ (and a corresponding Lorentz factor $\gamma$) relative to ${\cal I}'$ along the $z'$-direction. Then from a Lorentz transformation of the event ${\cal E}_1$, occurring at $t_1'=0,z_1'=z_0$ in ${\cal I}'$, we get 
\begin{eqnarray}
\label{eq:38.5a}
\nonumber
{t_1}&=& -z_0\gamma v/c^2 \\
{z_1}&=&  z_0\gamma\,
\end{eqnarray}
Thus according to observers in frame ${\cal I}$, the emission of the light ray will take place at a location $z_1=z_0\gamma$, at time ${t_1}= -z_0\gamma v/c^2$. Then, from (Eq.~(\ref{eq:38.5a})),  the distance travelled by the light ray between events ${\cal E}_1$ and ${\cal E}_2$, as seen in frame ${\cal I}$, is
\begin{eqnarray}
\label{eq:38.6a}
r=c\Delta t=c(t_2-t_1)=z_0\gamma v/c\,,
\end{eqnarray}

\begin{figure}[t]
\begin{center}
\includegraphics[width=\columnwidth]{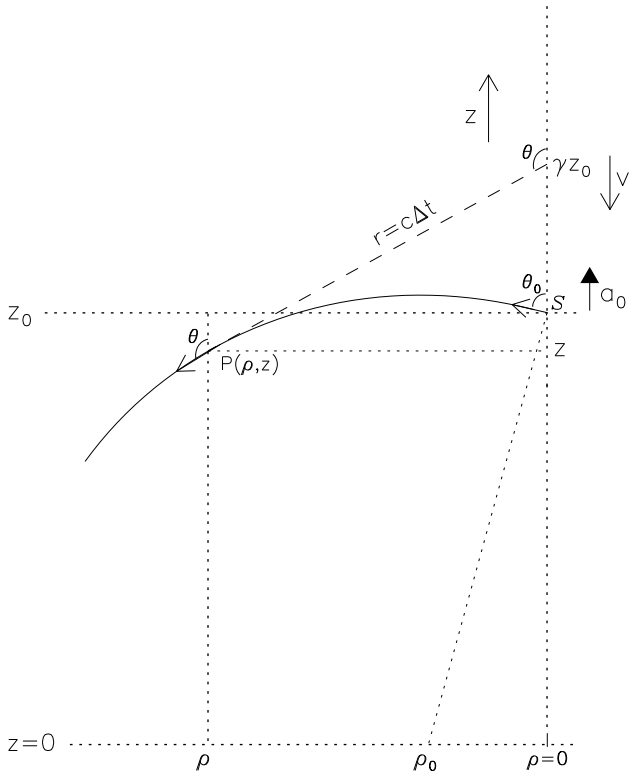}
\caption{Light initially emitted along direction $\theta_0$ from a source ${\cal S}$, stationary in the 'uniform' gravitational field at $z_0$, is later moving along $\theta$ at a point $P(\rho,z)$, the light-ray trajectory bent due to the gravity. By the equivalence principle the source ${\cal S}$ is stationary in the comoving accelerated frame ${\cal A}$. 
At the time of emission of the light, ${\cal I}'$ is the instantaneous inertial rest frame of the source ${\cal S}$. In another inertial frame ${\cal I}$, moving with a velocity $v$ relative to ${\cal I}'$, light moving along $\theta$ (dashed line) seems to be emitted at a time $t=-z_0\gamma v/c^2$ from the source ${\cal S}$ at $z=z_0\gamma$. 
The trajectory of the light ray (solid curve), continuously bent by gravity for an observer stationary in the equivalent gravitational field, turns out to be a circle of radius $(\rho_0^2+z_0^2)^{1/2}$, centered at $\rho_{0}$ in the $z=0$ plane (see text).}
\label{F6}
\end{center}
\end{figure}

Let $\theta$ be the angle along which the light ray moves in a straight line in inertial frame ${\cal I}$. To determine the relation between  $\theta$ and $\theta'$, we employ a Lorentz transformation between inertial frames ${\cal I}$ and ${\cal I}'$ for the space-time intervals between events ${\cal E}_1$ and ${\cal E}_2$, to get
\begin{eqnarray}
\label{eq:38.ab5}
\nonumber
{\Delta t}&=& \gamma(\Delta t' - v \Delta z'/c^2)\\
\nonumber
{\Delta z}&=&  {\gamma(\Delta z'- v\Delta t'})\\ 
\Delta \rho&=&\Delta \rho'
\,,
\end{eqnarray}
Noting that $\theta_0=\theta'$, from Eqs.~(\ref{eq:38.ab5a}) and (\ref{eq:38.ab5}) we get 
\begin{eqnarray}
\label{eq:38.ab6}
\nonumber
{\cos\theta}&=&  \frac{\Delta z}{c\Delta t}= \frac{{\cos\theta_0 - v/c}}{(1-{ v\cos\theta_0/c})}\\
{\sin\theta}&=&\frac{\Delta \rho}{c\Delta t}=  \frac{\sin\theta_0}{\gamma (1- v\cos\theta_0/c)}
\,.
\end{eqnarray}
Equation~(\ref{eq:38.ab6}) is actually the aberration formula \cite{RG70} where light emitted along $\theta_0$ in frame ${\cal I}'$ would appear to be moving along $\theta$ in frame ${\cal I}$, the latter having a velocity $v$ (and a corresponding Lorentz factor $\gamma$) with respect to frame ${\cal I}'$.
Since in frame  ${\cal I}'$, light rays are assumed to be emitted isotropically, half of the light rays will be at angles $\theta_0>\pi/2$, implying from Eq.~(\ref{eq:38.ab6}) that half of the total light rays in frame ${\cal I}$ 
will be moving within a cone of half-opening angle given by $\sin\theta=1/\gamma$ with respect to the $-z$ direction.

Since at events ${\cal E}_2$ the inertial frame ${\cal I}$ is coincident with ${\cal A}$, then from the clock and length hypotheses \cite{RI06}, all momentarily space-time measurements made by the accelerated observers in ${\cal A}$ will everywhere ($z>0$) match exactly with those made in the instantaneous rest frame ${\cal I}$. Thus in ${\cal A}$ too, at that instant,  light would appear to move along $\theta$. This is shown schematically in Fig.~\ref{F6} where the dashed line indicates the path of the light ray in frame ${\cal I}$ emitted along $\theta$ by the source ${\cal S}$ when it was at the location $z_1=\gamma z_0$. At point $P(\rho,z)$ the dashed line is tangent to the light-ray trajectory, shown as a solid curve, starting along $\theta_0$ from the source ${\cal S}$ position at $z_0$ in ${\cal A}$. Thus, as seen in the comoving accelerated frame ${\cal A}$, light emitted initially along $\theta_0$ at $z_0$, is later moving at $P(\rho,z)$ along $\theta$. Using the equivalence principle, we want to determine the trajectory of the light ray as it will appear to the stationary observers in the gravitational field, where the acceleration due to gravity is along the $-z$ direction.

From Fig.~\ref{F6}, we have the relations
\begin{eqnarray}
\label{eq:38.1}
z=\gamma z_0+r\cos\theta, \,\,\,\,\,\, \rho=r\sin\theta\,,
\end{eqnarray}
where $\theta$ is the angle with respect to the +ve $z$-axis. For $z<\gamma z_0$, $\theta>\pi/2$,  implying $\cos\theta<0$.

Then using (Eq.~(\ref{eq:38.6a})), we get
\begin{eqnarray}
\label{eq:38b2}
\frac{z}{z_0}=
\gamma+\frac{r\cos\theta}{z_0}=\gamma(1+v \cos\theta/c)\,.
\end{eqnarray}

It is interesting to note that a light ray emitted at ${z_0}$ in frame ${\cal A}$ in an `upward' direction, i.e. moving initially with $0<\theta_0<\pi/2$ when frame ${\cal I}'$ coincides with frame ${\cal A}$, will steadily lose its upward momentum due to the gravity and before its  velocity vector points in the `downward' direction, it will momentarily move in frame ${\cal A}$ along a horizontal direction ($\theta=\pi/2$) at a location ${z}=\gamma {z_0}$, which, from Eq.~(\ref{eq:38.ab6}), will be happening when an inertial frame moving with $v/c=\cos\theta_0$ or $\gamma=1/\sin\theta_0$ with respect to frame ${\cal I}'$, would be coincident with ${\cal A}$.

From Eq.~(\ref{eq:38.02}), a standard clock stationary at $z$, measures the proper time by a factor $z/z_0$ slower, as compared to another similar standard clock stationary at $z_0$. 
As a result, a light ray travelling from $z_0$ to $z$, will have its frequency, measured by observers using local clocks, shifted by $z_0/z=1/[\gamma(1+v \cos\theta/c)]$ (Eq.~(\ref{eq:38b2})). Thus as a light ray moves deeper in the gravitational field ($\theta>\pi/2$), its frequency will increase, while for a light ray 'climbing' in the gravitational field ($\theta<\pi/2$) its frequency would drop. This is the well-known phenomenon of the gravitational redshift \cite{MTW73,SC85,MO94,RI06,4}.

Now the inertial frame ${\cal I}'$ is the instantaneous rest frame of ${\cal S}$ at the time of emission of light, while the inertial frame ${\cal I}$ is moving with a velocity $v$ relative to ${\cal I}'$. Then $\delta=1/[\gamma(1+v \cos\theta/c)]$ is the kinematic relativistic Doppler factor by which the frequency of a light ray, as measured in inertial frame ${\cal I}$, will be higher than in the instantaneous rest frame ${\cal I}'$ \cite{RG70}, since in frame  ${\cal I}$ the source emits light along $\theta$ from  position $z=z_0\gamma$, while moving with an instantaneous velocity $v$ (Fig.~\ref{F6}). This is a kinematic explanation of the gravitational redshift in terms of the Doppler redshift, through the equivalence principle. 

In order to determine the light-ray trajectory, we first find its slope, ${{\rm d}\rho}/{{\rm d}z} =\tan\theta$, at a general point $P(\rho.z)$. 

From Eq.~(\ref{eq:38.1}), we have
\begin{eqnarray}
\label{eq:38a}
z_0^2\gamma^2+\rho^2=z^2+r^2-2 r z \cos \theta\,.
\end{eqnarray}
Also, using (Eq.~(\ref{eq:38.6a})), we can write  
\begin{eqnarray}
\label{eq:38ac}
z_0^2\gamma^2-r^2 =z_0^2\gamma^2[1-(v/c)^2] =z_0^2\,. 
\end{eqnarray}
Substituting in Eqs.~(\ref{eq:38a}), we get
\begin{eqnarray}
\label{eq:38a1}
2rz\cos \theta=z^{2}-z_0^{2}-\rho^{2}\,.
\end{eqnarray}
Then using $\rho=r\sin\theta$ (Eq.~(\ref{eq:38.1})), we get 
\begin{eqnarray}
\label{eq:38a2.3}
\tan \theta=\frac{2\rho z}{z^{2}-z_0^{2}-\rho^{2}}\,
\end{eqnarray}
Thus we get the slope of the light ray trajectory at point $P(\rho.z)$ (Fig.~\ref{F6}) as
\begin{eqnarray}
\label{eq:38.2a5.1}
\frac{{\rm d}\rho}{{\rm d}z} =\frac{2\rho z}{z^{2}-z_0^{2}-\rho^{2}}\,.
\end{eqnarray}
It has a solution, $(\rho-\rho_{0})^{2}+z^{2}=\rho_{0}^{2}+z_0^{2}$, as can be verified easily by a substitution. The light ray thus follows a trajectory circular in shape with center at $(\rho=\rho_{0},z=0)$ and a radius $(\rho_0^2+z_0^2)^{1/2}$, thus passing through the source position at ($\rho =0,z=z_0)$ (Fig.~\ref{F6}).

\begin{figure}[t]
\begin{center}
\includegraphics[width=\columnwidth]{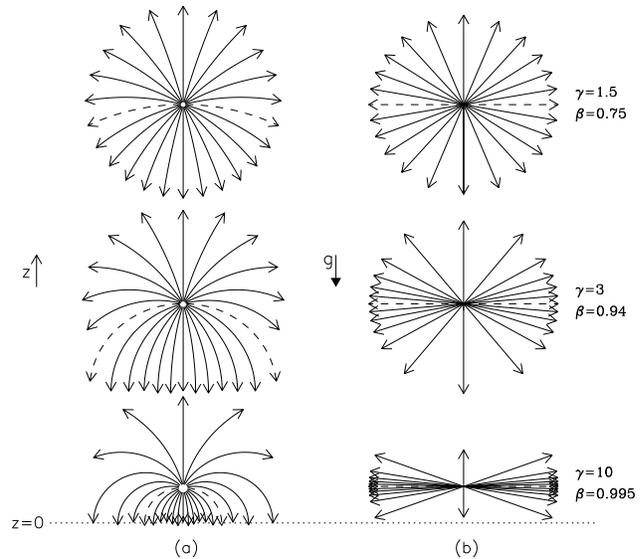}
\caption{A schematic representations of light rays originating from a source (a) stationary with respect to the local observers, supported at different depths in a `uniform' gravitational field (b) freely falling at different instants, as it gathers speed with respect to the observers stationary in the gravitational field. The dashed lines represent light rays that start from the source in the horizontal plane, i.e. in a plane normal to the $z$-direction.}
\label{F7}
\end{center}
\end{figure}

Figure~\ref{F7}a shows the trajectories of light rays originating from a source stationary with respect to the local observers, supported at different depths in a `uniform' gravitational field. 
It appears that the light-ray trajectories starting radially from the source position 
get continuously `bent' by the gravitational field, becoming vertical as one approaches the horizon, $z=0$. In the comoving accelerated frame ${\cal A}$, light rays arriving at $z\rightarrow0$ at $t=0$, would have been emitted in the corresponding inertial frame at $z\rightarrow \infty$ with $\gamma\rightarrow \infty$, and will appear to be moving perpendicular to the $z=0$ plane. 
Actually for any initial $\theta_0$, from  Eq.~(\ref{eq:38.ab6}),  $\theta\rightarrow \pi$ as $\gamma\rightarrow \infty$; light rays will thus be moving parallel to the $z$-axis along $-z$ direction as they approach the horizon, $z=0$. 

\subsection{Light rays being emitted from a source {\em falling freely} in the gravitational field}\label{S2B}
Figure~\ref{F7}b shows the trajectories of light rays originating from a source freely falling at different instants, as it gathers speed with respect to the observers stationary in the gravitational field, where light-ray trajectories, extending along radial straight lines from the instantaneous position of the source, are bunched about the horizontal plane, normal to the direction of motion, passing through the instantaneous position of the source that during its free fall picks up speed because of acceleration due to gravity.
Even though the trajectories of individual light rays after being emitted from a freely falling source, as seen by  observers stationed in the comoving accelerated frame or in the equivalent  gravitational field, may appear to be bent with respect to the time-retarded position of the emitting source, nevertheless, the same observers will find the light rays to be spreading in radial straight lines with respect to the instantaneous position of the freely falling source.

A source freely falling in a uniform gravitational field is actually stationary in a co-falling  inertial frame, say ${\cal I}$, and the light rays emitted say, isotropically, continue to move in  initial radial directions from the stationary charge position in  ${\cal I}$. However, as the source due to its free fall gathers speed along with ${\cal I}$, to the observer stationary in the gravitational field there is a Lorentz contraction of distances between horizontal planes of  ${\cal I}$. Therefore, as seen by the observer stationary in the gravitational field, there is a bunching of the light-ray trajectories toward the horizontal plane passing through the instantaneous position of the freely falling source. As the  freely falling source nears $z=0$ plane, the bunching of the light-ray trajectories gets so exaggerated that almost all light rays lie within a narrow angle $\sim 1/\gamma$ about the horizontal plane where $\gamma$ is the instantaneous Lorentz factor of motion of the freely falling source, as seen in Fig.~\ref{F7}b.

Here a rather puzzling question could arise. It is well-known that for a relativistically moving source, light rays, due to aberration, appear to get concentrated mostly in a narrow cone of opening angle $\sim 1/\gamma$ in the {\em forward} direction, an effect known as relativistic beaming \cite{RG70}. The forward direction of motion here is downward, along $-z$. Then, does it not contradict the distribution of light ray within a narrow angle $\sim 1/\gamma$, but bunched rather about the horizontal plane, the plane that is {\em normal} to the direction of motion, as is seen in Fig.~\ref{F7}b? This bunching of light rays about the horizontal plane, instead of beamed in the forward direction, is discussed in detail in \ref{SAC}. 
\section{Electric field lines versus the trajectories of light rays emitted from a source at the charge position in a gravitational field}\label{S3}
The electric field in the case of a uniformly accelerated charge, in its instantaneous rest-frame, can be written in terms of the {\em retarded-time} position of the charge as \cite{88}
\begin{eqnarray}
\label{eq:38.abb}
{\bf E}=\left[\frac{e{\bf n}}{\gamma ^2 r^2(1-{\bf n}\cdot{\bf v}/c)^2}\right]_{t_{\rm r}}=\left[\delta^2\frac{e{\bf n}}{r^2}\right]_{t_{\rm r}}\,,
\end{eqnarray}
where $\delta=[\gamma(1-{\bf n}\cdot{\bf v}/c)]^{-1}$ is the Doppler factor. It can be seen that the electrostatic field lines in this case seem to emanate from the charge like light rays, emitted radially outward from a source at the  retarded-time charge position. Not only does the field strength, like light intensity, fall $\propto 1/r^2$ with respect to the source position at the retarded time $r/c$ earlier, 
since the electric field influence propagates with a speed $c$, even 
the field strength is $\propto \delta ^2$, in the same way as the light-ray trajectories, due to aberration of light, would be concentrated in the forward direction of motion by a factor $\delta ^2$, when considered with respect to the time-retarded position of the source \cite{RI06,RG70}.

The electric field of a uniformly accelerated charge (Eq.~(\ref{eq:38.abb})), momentarily stationary in the inertial frame, say ${\cal I}$, can be expressed alternatively in terms of the `present' ({\em real-time}) position of the above charge and is given by Eq.~(\ref{eq:38.ab1}) \cite{88}.
Then, with respect to the charge position in the comoving accelerated frame ${\cal A}$, having ${\cal I}$ its instantaneous coincident frame, or by the equivalence principle, for a charge supported in a gravitational field, it is possible to relate the continuously changing direction $\theta$ of the electric field vector to its initial starting direction $\theta_0$, which is radial from the instantaneous charge position at $z_0$ (Fig.~\ref{F1}).

Accordingly, from Eq.~(\ref{eq:38.ab1}), the slope of an electric field line in frame ${\cal A}$, or equivalently in the gravitational field, is
\begin{eqnarray}
\label{eq:38.2a5}
\tan\theta=\frac{{\rm d}\rho}{{\rm d}z} = \frac{E_{\rho}}{E_{\rm z}} =\frac{2\rho z}{z^{2}-z_0^{2}-\rho^{2}}\,.
\end{eqnarray}
A comparison with Eq.~(\ref{eq:38.2a5.1}) shows the slope of the electric field lines to be exactly the same as of light-ray trajectories in the gravitational field. Whether it is an electric field line or it is a light ray, emanating from a corresponding source stationary in a gravitational field, due to the gravitational bending, either of the two traces a curved path $(\rho-\rho_{0})^{2}+z^{2}=\rho_{0}^{2}+z_0^{2}$ (Eq.~(\ref{eq:38.2a4})), which, starting from the source position at ($\rho =0,z=z_0)$, describes a circle with center at $(\rho=\rho_{0},z=0)$ and a radius $(\rho_0^2+z_0^2)^{1/2}$, (Figs.~\ref{F1} and~\ref{F6}).

Thus, a comparison of (Fig.~\ref{F4}a) and (Fig.~\ref{F7}a) shows that the electric field lines of a charge, supported in a gravitational field exactly follow the trajectories of light rays emitted in all directions from a source at the charge position. As the light-ray trajectories bend in the gravitational field, so do the electric field lines. 

On the other hand, the electric field configuration for a freely falling charge, shown in Fig.~\ref{F4}b, are always in radially straight lines from the instantaneous position of the charge, but bunched toward the horizontal plane normal to the direction of free fall, and so do follow the trajectories of light rays spreading away from a source freely falling in the gravitation field (Fig.~\ref{F7}b).

Thus the electric field lines of a charge, whether supported or freely falling in a gravitational field, follow exactly the trajectories of light rays emitted isotropically from a source at the corresponding charge location.

\section{Flux through horizontal planes ``below'' or ``above'' the source in gravitational field}\label{S4}
We shall examine here whether due to gravitational bending of electrostatic field lines or of  light-ray trajectories, there are any changes in the total electric flux as well as the flux of light ray passing through a horizontal plane ``below'' or ``above'' the source, stationary in the gravitational field. For that, we compute the flux through a horizontal plane, say $P_1$ at $z=z_1$, that lies below the source as well as for a plane, say $P_2$ at $z=z_2$, above the source. For a comparison, we shall also calculate the respective flux through similar horizontal planes for a source freely falling in the gravitational field. 
\subsection{Electric flux crossing a horizontal plane ``above'' or ``below''}\label{S4A}

\begin{figure}[t]
\begin{center}
\includegraphics[width=\columnwidth]{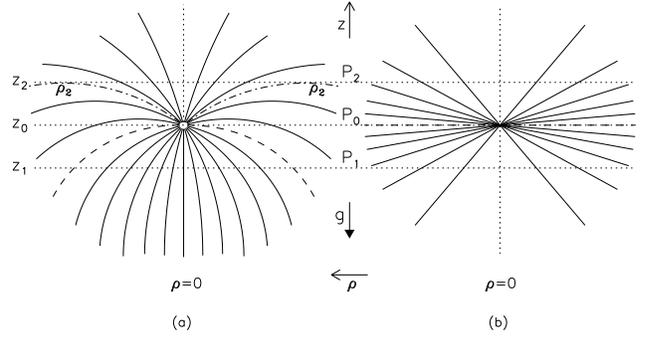}
\caption{Electric flux computation for a charge, at $z_0$ in the gravity field. Shown are two planes, $P_1$ and $P_2$, normal to the $z$-direction, below and above the charge at $z_1<z_0$ and $z_2>z_0$ respectively, used for calculating the electric flux due to the charge through each of them. Also shown is a plane $P_0$, that passes through the charge at $z=z_0$. Field lines start isotropically in radial directions from the charge. The dashed lines represent the field lines starting from the charge in the horizontal plane $P_0$, i.e. in a direction normal to the $z$-axis.
(a) For a charge stationary in the gravitational field, the electric field lines, though starting  radially from the charge, get bent downward into circular shapes due to gravity. The dash-dotted lines represent the  gravitational bent field lines that touch the plane $P_2$ tangentially at $\rho_2$. (b) For a freely falling charge in the gravitational field, electric field lines extend in radial directions from the instantaneous position of the charge, albeit bunched around the horizontal plane $P_0$, passing through the charge.}
\label{F9}
\end{center}
\end{figure}

For a charge {\em stationary} in the gravitational field (Fig.~\ref{F9}a), electric flux $F_1$, through the plane $P_1$ lying at $z_1$ below the charge at $z_0$, using Eq.~(\ref{eq:38.ab1}), is
\begin{eqnarray}
\label{eq:38.3a}
F_1&=&\int_{0}^{\rho}E_{\rm z} 2\pi\rho\,{\rm d}\rho\nonumber\\
&=&\!\!4e\!\int_{0}^{\rho}\!\frac{-z_0^{2}(z_0^{2}-z_1^{2}+\rho^{2})}{[(z_0^{2}-z_1^{2}-\rho^{2})^{2}+4z_0^{2}\rho^{2}]^{3/2}} 2\pi\rho\,{\rm d}\rho.
\end{eqnarray}

With the help of the expression
\begin{eqnarray}
\label{eq:38.4}
\frac{{\rm d}}{{\rm d}\rho}\frac{z_0^{2}-z^{2}-\rho^{2}}{[[(z_0^{2}-z^{2}-\rho^{2})^{2}+4z_0^{2}\rho^{2}]^{1/2}}\nonumber\\
&&\!\!\!\!\!\!\!\!\!\!\!\!\!\!\!\!\!\!\!\!\!\!\!\!\!\!\!\!\!\!\!\!\!\!\!\!\!\!\!\!\!\!\!\!\!\!\!\!\!\!\!=\frac{-2\rho z_0^{2}(z_0^{2}-z^{2}+\rho^{2})}{[[(z_0^{2}-z^{2}-\rho^{2})^{2}+4z_0^{2}\rho^{2}]^{3/2}},
\end{eqnarray}
we get 
\begin{eqnarray}
\label{eq:38.8}
F_1&=&2\pi e \Big|\frac{z_0^{2}-z_1^{2}-\rho^{2}}{[[(z_0^{2}-z_1^{2}-\rho^{2})^{2}+4z_0^{2}\rho^{2}]^{1/2}}\Big|_{0}^\rho\,.
\end{eqnarray}
It can be easily verified that an electric flux $2\pi e$ along $-z$ direction passes through the plane $P_1$ from  $\rho=0$ to an upper limit $\rho_1=\sqrt{z_0^{2}-z_1^{2}}$, while an equal flux  $2\pi e$ along $-z$ direction passes through the plane $P_1$ from $\rho_1$ to $\rho \rightarrow \infty$. 

Thus the total flux through an infinite plane at $z_1$, from $\rho=0$ to $\rho \rightarrow \infty$, is 
\begin{eqnarray}
\label{eq:38.9}
F_1=2\pi e\left[-1-\frac{z_0^{2}-z_1^{2}}{z_0^{2}-z_1^{2}}\right]={-4\pi e }\,,
\end{eqnarray}
the negative sign indicating that the electric flux through the plane $P_1$ is in downward direction.

In the same way we can compute the electric flux $F_2$ through the plane $P_2$ at $z=z_2$, that lies above the charge in the gravitational field (Fig.~\ref{F9}a), to get
\begin{eqnarray}
\label{eq:38.3}
F_2 =4e\int_{0}^{\rho}\frac{z_0^{2}(z_2^{2}-z_0^{2}-\rho^{2})}{[[(z_2^{2}-z_0^{2}-\rho^{2})^{2}+4z_2^{2}\rho^{2}]^{3/2}} 2\pi\rho\,{\rm d}\rho\,.
\end{eqnarray}
Here the outward normal to the plane $P_2$ is along $z$ direction  with $z_2 > z_0$. From Eq.~(\ref{eq:38.3}) it can be seen that for $\rho<\rho_2$, where $\rho_2=\sqrt{z_2^{2}-z_0^{2}}$, the electric field is along $+z$ direction, making a +ve contribution to the electric flux through the  plane $P_2$, while for $\rho>\rho_2$, the electric field is along $-z$ direction, making a -ve contribution to the electric flux through the  plane $P_2$. 

Again, using Eq.~(\ref{eq:38.4}), we get
\begin{eqnarray}
\label{eq:38.5}
F_2&=&-2\pi e \Big|\frac{z_2^{2}-z_0^{2}+\rho^{2}}{[[(z_2^{2}-z_0^{2}-\rho^{2})^{2}+4z_2^{2}\rho^{2}]^{1/2}}\Big|_{0}^\rho\,.
\end{eqnarray}

Fig.~\ref{F9}a shows a pair of (dash-dotted) field lines, starting radially from the charge at $z_0$, but getting bent due to gravity in circular shapes so as to touch the plane $P_2$ tangentially at $\rho_2$. It is clear that these represent in limit the last field lines that pass through $P_2$ along $+z$ direction, intersecting the plane between  $\rho=0$ and $\rho=\rho_2$. An evaluation of electric flux passing through $P_2$ between $\rho=0$ and $\rho_2=\sqrt{z_2^{2}-z_0^{2}}$ yields $2\pi e(1-\rho_2/z_2)$
along $+z$ direction. Also, an equal amount of flux along $-z$ direction passes through the plane $P_2$, from $\rho=\rho_2$ to $\rho \rightarrow \infty$, making net flux  nil. 
Thus the total flux through an infinite plane at $z_2$, from $\rho=0$ to $\rho \rightarrow \infty$, is 
\begin{eqnarray}
\label{eq:38.7}
F_2=-2\pi e\left[1-\frac{z_2^{2}-z_0^{2}}{z_2^{2}-z_0^{2}}\right]=0\,.
\end{eqnarray}

One can visualize as well as compute the limiting radius $\rho_2$ on the plane $P_2$ for the field lines that pass through $P_2$ along $+z$ direction and the electric flux contained within them, in another way. In Fig.~\ref{F1}, if we choose $\rho_0$ on the $z=0$ plane at $\sqrt{z_2^{2}-z_0^{2}}$, then the vertical straight line at $\rho_0$, normal to the $z=0$ plane, intercepting the plane $P_2$ at $\rho_2=\rho_0=\sqrt{z_2^{2}-z_0^{2}}$, will be normal to the plane $P_2$  at $ z_2$ as well. 
Then the circular field line with center at $\rho_0=\sqrt{z_2^{2}-z_0^{2}}$ on the $z=0$ plane, and a radius $\sqrt{\rho_0^{2}+z_0^{2}}=z_2$, will be touching the plane $P_2$ tangentially at $\rho_2$, as one of the dash-dotted lines shown in Fig.~\ref{F9}a.
The corresponding starting angle $\theta_0$ (Fig.~\ref{F1}) for this field line would be given by $\cos \theta_0=\rho_0/\sqrt{\rho_0^{2}+z_0^{2}}=\rho_2/z_2$. Thus the electric flux passing through $P_2$, along $+z$ direction, between $\rho=0$ and $\rho_2$ is $2\pi e(1-\rho_2/z_2)=2\pi e(1-\cos \theta_0)$.
This, of course, is the expected value of the electric flux within a solid angle $2\pi (1-\cos \theta_0)$, for the field lines starting from the charge initially isotropically. 

From Fig.~\ref{F9}a it is clear that due to gravitational bending, the upward pointing electric field lines spread out in the horizontal plane, as a result the electric flux in the upward direction is less than in the downward direction. In other words, due to the charge at $z_0$, the  net electric flux through a horizontal circular surface of a given radius $\rho$, held at $z_2>z_0$ is less than the flux through a similar surface at $z_1<z_0$, of course assuming $z_2-z_0=z_0-z_1$. It also seems that net electric flux through the infinite plane $P_1$ ($P_2$) is independent of its distance below (above) the charge, and that through each horizontal, infinite plane there is an extra net flux of an amount $-2\pi e$, than what would be without gravity. The negative sign indicates that the extra flux is in the downward direction, i.e. along the direction of gravity. It would be interesting to check for the plane $P_0$, containing the charge at $z=z_0$ (Fig.~\ref{F9}a), what might be the electric flux, which otherwise, in the absence of gravity, is supposed to be nil. Then putting $z_1=z_0$ in Eq.~(\ref{eq:38.8}), or $z_2=z_0$ in Eq.~(\ref{eq:38.5}), we have
\begin{eqnarray}
\label{eq:38.10}
F_0&=&-2\pi e \Big|\frac{\rho}{[\rho^{2}+4z_0^{2}]^{1/2}}\Big|_{0}^\rho
\,,
\end{eqnarray}
which for $\rho \rightarrow \infty$ yields 
\begin{eqnarray}
\label{eq:38.12}
F_0=-2\pi e\,.
\end{eqnarray}
The negative sign indicates that the electric flux through the plane $P_0$ is downwards, i.e. in the $-z$ direction. 

Thus we find that the total electric flux in a vertical direction in a gravitational field differs from the value expected in the absence of the gravitational field,  through each infinite horizontal plane by $-2\pi e$. 

For a comparison, we can also compute the electric flux for a charge {\em freely falling} in the gravitational field. For that, we can exploit Eq.~(\ref{eq:38.03}) to get the electric flux through $P_1$ as (Fig.~\ref{F9}b)
\begin{eqnarray}
\label{eq:38.04}
\int_{0}^{\infty}E_{\rm z} 2\pi\rho\,{\rm d}\rho &=&-2\pi e\int_{0}^{\infty} \frac{\gamma \Delta z}{[\gamma^{2}(\Delta z)^{2}+\rho^{2}]^{3/2}} \rho\,{\rm d}\rho\nonumber\\
&=&-2\pi e\,,
\end{eqnarray}
where we have used
\begin{eqnarray}
\label{eq:38.05}
\frac{{\rm d}}{{\rm d}\rho}\frac{1}{[\gamma^{2}(\Delta z)^{2}+\rho^{2}]^{1/2}}=\frac{-\rho}{[\gamma^{2}(\Delta z)^{2}+\rho^{2}]^{3/2}}\,.
\end{eqnarray}
The computed flux, $2\pi e$ along $-z$ direction, through $P_1$, is exactly as one would expect from Gauss law. A similar calculation of the electric flux through the infinite plane $P_2$ at $z=z_2$, above the charge, yields $2\pi e$, which is along $+z$ direction, again a result consistent with Gauss law in the absence of gravity.

The number of field lines crossing small hemispheres, in the vicinity of the charge, where the field lines are still in radial directions from the charge, seem to be still $2\pi e$ in outward   direction, the value expected from Gauss law, and if the field lines had continued in their original directions, the flux through a plane like $P_2$ would be $2\pi e$. However, these upward pointing field lines not only spread out due to gravitational bending, sooner or later they turn fully to point in the downward direction, thereby cancelling the net upward flux through $P_2$. But what is surprising is that the fact that the electric flux is zero through $P_2$ or that the electric flux is $4\pi e$ through $P_1$, does not depend upon the strength of gravity at the charge location, or even on the value of $g$ in the respective plane. Does it mean even on Earth's gravity,  $g \approx 9.8$ m sec$^{-2}$, one should expect to find zero net electric flux through a horizontal plane above a charge supported against gravity in Earth's gravitational field? Actually that would indeed be the case if in the gravitational field under investigation there existed a plane that extended to infinity with the same gravitational field everywhere, so that due to gravitational bending, all field lines going through the plane in an upward direction would ultimately cross the plane in a downward direction as well, making the net flux zero.
Actual measurements of electric flux through a finite spherical cross-section around Earth may not easily reveal departures from what calculated from Gauss law in the absence of gravity.

We may like to get an idea of the distance $\rho$ in a horizontal plane, say, at $z=z_0$, beyond which there may be an appreciable number of field lines crossing in the downward direction due to the gravitational bending, giving rise to a departure from what expected otherwise in the absence of the gravitational field. For this, we note from Eq.~(\ref{eq:38.10}) that up to a finite $\rho$, electric flux through the plane $P_0$ is
\begin{eqnarray}
\label{eq:38.13}
F_0= \frac{-2\pi e\rho}{[\rho^{2}+4z_0^{2}]^{1/2}}\,.
\end{eqnarray}
For $\rho\ll z_0=c^2/g$, $F_0\approx -\pi e \rho g/c^2$. Thus for $g \approx 9.8$ m sec$^{-2}$, the acceleration due to gravity on earth, $c^2/g \approx 10^{16}$m, about one light year. Then for $\rho\sim 2$m, the amount of electric flux that would pass through is only about one part in $10^{16}$ or so of what would be the case for an infinite plane, supported at $z_0$ in a constant $g$. 
Only for $\rho\sim z_0$, we get an appreciable value of $F_0$ that departs from what expected from Gauss law in the absence of gravity.  

An interesting consequence of the gravitational bending of electric field lines is for an infinite  sheet with a uniform surface charge density $\sigma$, held in a horizontal direction in the gravitational field, say at $z_0$. Due to the effect of gravity, there will be no electric field in the regions `above' the sheet (at $z>z_0$), as the net electric flux contribution of each charge element of the infinite sheet is zero, while `below' the sheet (at $z<z_0$) there will be a uniform electric field of strength  $4\pi \sigma$ in the downward direction. Through the sheet (at $z=z_0$), the field will be  $2\pi \sigma$ in the downward direction. This implies a gravity-induced self-force  $2\pi \sigma^2$ per unit area on the charged sheet in a gravitational field. Surprisingly, though it is  in the downward direction, along the direction of gravity, but it is independent of the strength $g$ of gravity, seemingly due to  its infinite extent being immersed in a constant $g$. One can compare it with a small spherical shell of uniform surface charge density $\sigma$, where on each surface element of the shell, there is an outward repulsive force, $2 \pi \sigma ^{2}$ per unit area  \cite{PU85,Si92}, which due to the spherical symmetry, becomes zero when integrated on the whole shell. But there is a finite gravity-induced net force $2Q^2{\bf g_0}/3r_0 c^2$ (Eq.~(\ref{eq:38a.4})) directly proportional to ${\bf g_0}$, the acceleration due to gravity, on the shell with a  total charge $Q$.

For a hypothetical parallel plate capacitor of infinite plate dimensions with uniform surface charge density $\pm \sigma$, with plate separations along the $z$ direction, the field in the region between the capacitor plates will be still $4\pi \sigma$ and no electric field above or below the capacitor plates, as would be the case without gravity. Even the effect of the gravity-induced self-force does not show up as an extra force. The upper plate feels no electric field due to the lower plate but has its own self-force  $2\pi \sigma^2$ per unit area in the downward direction, while the lower plate has an upward force $4\pi \sigma^2$ per unit area due to the upper plate which gets reduced by its own self-force  $2\pi \sigma^2$ per unit area in the downward direction thus leaving an upward force $2\pi \sigma^2$ per unit area. The capacitor plates feel the same  force, $2\pi \sigma^2$ per unit area, as would be the case, in the absence of gravity, due to the electrical attraction between the plates \cite{PU85,Si92}. 
\subsection{Light-ray trajectories crossing the horizontal planes ``above'' or ``below''}\label{S4B}

Similar conclusions can be drawn for the trajectories of light rays emitted from a source in the gravitational field. For example, we can compute the flux of light-rays crossing the Plane $P_2$ in upward direction. Let $n$ be the number of light rays emitted per unit solid angle by the source, where each ray is associated with a fixed amount of flux. From Eq.~(\ref{eq:38a1}), the last light ray trajectory encountering the Plane $P_2$ at $z_2$ is for $\rho_2=\sqrt{z_2^{2}-z_0^{2}}$, where $\theta=\pi/2$, implying that the light ray trajectory becomes horizontal and is just touching the Plane $P_2$. If $\theta_0$ is the corresponding initial angle of emission for this light ray trajectory, then the number of light-ray trajectories crossing $P_2$, along $+z$ direction, between $\rho=0$ and $\rho_2$ is $2\pi n(1-\cos \theta_0)$, assuming an isotropic initial distribution.
Now from Eq.~(\ref{eq:38.ab6}), $\cos\theta=0$ implies $\cos\theta_0=v/c=\sqrt{1-1/\gamma^2}$, where we also have from Eq.~(\ref{eq:38.1}), $z_2=\gamma z_0$, thereby giving us $\cos\theta_0=\sqrt{1-z_0^2/z_2^2}=\rho_2/z_2$, which implies $2\pi n(1-\rho_2/z_2)$ as the number of light-ray trajectories crossing upward through $P_2$.  The same light-ray trajectories would bend downward due to gravity, passing through the plane $P_2$ in $-z$ direction, between $\rho=\rho_2$ to $\rho \rightarrow \infty$, making the net light ray flux through the plane $P_2$ nil.
Thus the proportional number of light-ray trajectories crossing through $P_2$, whether upward or downward, is the same as that for electric field lines.

Actually this happens because, as we have shown earlier, light-ray trajectories are influenced by the gravitational field exactly the same way as the electric field lines. 
Thus for an infinite horizontal sheet (a plane, fully transparent to light rays) held above a source {\em stationary} in the gravitational field, the net flux of light ray numbers crossing the sheet will be zero, while for a similar horizontal sheet below the source, the net flux will comprise all light rays emitted from the source as the gravitationally bent trajectories of all these light rays would be passing through the sheet in the downward direction. Of course, for a freely falling source in the gravitational field, the net flux of light ray through a horizontal plane will be the same as in the absence of gravity. Thus our conclusions are as much applicable to electric field lines, whether  for a charge stationary (Fig.~\ref{F4}a) or freely falling (Fig.~\ref{F4}b), as to the trajectories of light rays from a source stationary (Fig.~\ref{F7}a) or freely falling (Fig.~\ref{F7}b). 
\section{In the vicinity of the event horizon}\label{S5}
Although the spatial extent of the accelerated frame ${\cal A}$ is restricted to $z>0$, however,  the coincident inertial frames do extend from $-\infty$ to $+\infty$ along the $z$ direction (see section~\ref{S0B}). Since signals from a source of light rays (or an electric charge) located at $z\le 0$ in an inertial frame ${\cal I}$, could be seen to reach a location $z>0$ at time $t=0$, when  ${\cal I}$ coincides with  ${\cal A}$, a question could be raised whether arrival of such signals would not be seen by observers stationary in ${\cal A}$. Actually, as seen by a uniformly accelerated observer, 
such signals must have been emitted in the infinite past in the observer's time \cite{RI06,10}, One can understand from the point of view of the observer stationary in the gravitational field that a signal from $z=0$ itself will need an infinite time to reach the observer. Effectively a signal, whether a light ray or an electric field, from a source at or beyond the horizon ($z\le 0$) will not reach the observer in any finite amount of time.

In the instantaneous rest frame ${\cal I}$ of the uniformly accelerated charge, as one approaches $z\rightarrow 0$, the electrostatic field, from Eq.~(\ref{eq:38.ab1}), turns perpendicular to the $z=0$ plane, with a finite value  
\begin{eqnarray}
\label{eq:38.3c}
E_{\rm z}\Big|_{z=0+} = \frac{-4 e z_0^{2}}{(z_0^{2}+\rho^{2})^2}
\end{eqnarray}\,.

Since at $t=0$, the situation under consideration, the electromagnetic field is absent at $z\le0$, it means a discontinuity in the electric field at $z=0$ plane, implying an incompatibility with Gauss law. 

To restore the compatibility with Gauss law, it has been said \cite{MS85} that there may be a finite surface charge density at the $z=0$ plane.
\begin{eqnarray}
\label{eq:38.4b}
\sigma=\frac{1}{4\pi} \nabla \cdot {\bf E}=\left.\frac{E_z}{4\pi} \right|_{\rm z=0} =\frac{-e z_0^{2}}{\pi (z_0^{2}+\rho^{2})^2}\,
\end{eqnarray}
which when integrated over the $z=0$ plane amounts to a total charge $-e$.

But no such charge distribution at the $z=0$ plane is pre-assigned in the case of a uniformly accelerated charge. Instead, consistency with Maxwell's equations is restored if one assumes that prior to the onset of acceleration at say, $t=-\tau$, the charge moved with a uniform velocity $v_{\rm i}$. Then in the limit $\tau \rightarrow \infty$, $v_{\rm i} \rightarrow c$ with the corresponding Lorentz factor $\gamma_{\rm i} \rightarrow \infty$, then one obtains at the $z=0$ plane a $\delta$-field \cite{3,JF15,88}
\begin{eqnarray}
\label{eq:38.4a}
E_\rho&=& \frac{2e\rho}{z_0^{2}+\rho^2}\delta(z)\,,
\end{eqnarray}
Now, if one takes the delta field (Eq.~(\ref{eq:38.4a})) at the $z=0$ plane into account, then one gets another contribution to $\nabla \cdot {\bf E}$
\begin{eqnarray}
\label{eq:38.4c}
\left.\frac{1}{\rho}\frac{\partial (\rho E_{\rho})} {\partial \rho}\right|_{\rm z=0}=\frac{4e z_0^{2}}{(z_0^{2}+\rho^{2})^2}\delta(z)\,
\end{eqnarray}
which makes the  $\nabla \cdot {\bf E}=0$ at $z=0$, thus restoring the compatibility with Gauss law as well as removing the necessity of assigning a surface charge density at the $z=0$ plane.  

However, since by equivalence principle, a charge permanently stationary at $z_0$ in the gravitational field is equivalent to a charge uniformly accelerated {\em for ever}, it may not be appropriate here to consider the field arising from a uniform motion of the charge before an onset of the acceleration, in a limiting case at some infinite past. Even otherwise, as we have demonstrated earlier, the electric field lines in a gravitation field follow the trajectories of light rays emitted from the charge position, there is no reason for the light-ray trajectories to suddenly switch near the horizon from a freely falling vertical direction into a horizontal direction. Actually as one approaches the horizon ($z\rightarrow 0$), from Eq.~(\ref{eq:38.02}), the passage of proper time is slower by a factor $z/z_0$, as compared to the proper time at the source location at $z_0$, which in our convention is also the coordinate time. Thus as far as the observers in frame ${\cal A}$ are concerned, a light ray emitted from a source at a finite $z$, e.g. $z_0$, would never reach the $z=0$ plane as it would require an infinite time of the observer to reach there. Thus even the electric field lines will remain confined to the $z>0$ region only.

In fact as one approaches  the horizon i.e. $z\rightarrow 0$, from Eq.~(\ref{eq:38.2a5}), $\theta\rightarrow \pi$ and, irrespective of the initial radial directions ($\theta_0$) of the field lines from the charge, all electric field lines point vertically downward (Fig.~\ref{F4}a), as also is the case for the light rays (Eq.~(\ref{eq:38a2.3})), which too, are moving parallel to the negative $z$ direction,  close to the horizon at $z=0$ (Fig.~\ref{F7}a).


However, if we shift the stationary position of the charge closer to $z=0$ plane, by choosing $z_0$ smaller, and consequently $g_0=c^2/z_0$ larger, then bending of electric field lines is  more extreme, and the electric flux through $P_1$ increases even for relatively smaller $\rho$ values. At the same time upward electric flux through $P_2$ reduces rapidly as field lines spread out horizontally for even small $\rho$ values, when the charge location is moved near to the $z=0$ plane. As $z_0$ approaches the horizon, almost all of the flux passes through a small patch of $P_1$ around the $z$-axis, towards the horizon and any electric field in regions above the charge will be reducing considerably.

To get a quantitative idea, from Eq.~(\ref{eq:38.8}), we find the electric flux passing downward through a circle of radius $\rho$ in the plane $P_1$ at $z_1$, for $z_1$ near the event horizon, i.e. for $z_1 \rightarrow 0$, to be $4\pi e {\rho^{2}}/(z_0^{2}+\rho^{2})$.
From this we see that the total electric flux, i.e. $2\pi e$, that would have otherwise passed through the whole plane $P_1$ for a charge outside the gravitational field, in the case of a charge stationary at $z_0$ in the gravitational field, passes through a circle of radius $\rho=z_0$ at or near the event horizon. Of course, when $z_0$ is chosen close to the horizon, i.e. $z_0 \rightarrow 0$, then the electric flux also passes through a proportionally smaller circle.

It was earlier shown that in the case of light rays, for a source moving with a velocity $v$ in an inertial frame, half of the emitted light rays lie within a cone of half-opening angle, given by $\sin\theta=1/\gamma$, around the direction of motion of the source. That means half of the light rays lie within a circle of radius $\rho=r\sin\theta={r}/{\gamma}$, on a horizontal plane at $z$, which, from Eq.~(\ref{eq:38.6a}), is calculated to be 
\begin{eqnarray}
\label{eq:38.3d}
\rho=\frac{r}{\gamma}=\frac{z_0 v}{c}\,.
\end{eqnarray}

Now in the inertial frame ${\cal I}$, instantaneously coincident with ${\cal A}$ at $t=0$, light rays approaching the horizon (at $z=0$), were emitted by the source at time $t\rightarrow -\infty$  when the source was at $z \rightarrow \infty$, moving along $-z$ direction with $v\rightarrow c$ (Fig.~\ref{F6}). Out of these one half of the light rays in the $z=0$ plane, from Eq.~(\ref{eq:38.3d}), would have trajectories lying within a circle of radius $\rho=z_0$. 

One can understand it in another way. From Fig.~\ref{F6}, the circular light-ray trajectories starting horizontally from the source at $z_0$, i.e. along $\theta_0=\pi/2$, would have the center of the circle at $\rho_0=0$ and a radius  $z_0$, intersecting the $z=0$ plane at $\rho=z_0$. Thus all those light rays that start with $\theta_0>\pi/2$, that is half of the total light rays emitted by the source, would be lying within the bounding circular light ray trajectory  $\theta_0=\pi/2$, and would reach the $z=0$ plane within a circle of radius $\rho=z_0$. This is exactly like the electric field lines from a charge stationary at $z_0$ in the gravitational field, where half of the field lines pass through a circle of radius $\rho=z_0$ in the $z=0$ plane. 
If location of the charge or light ray source is chosen in the vicinity of the event horizon, i.e. $z_0 \rightarrow 0$, then the region through which most of the electric flux or light-ray trajectories pass through in the $z=0$ plane, gets reduced to smaller and smaller values (Figs.~\ref{F4}a or ~\ref{F7}a).
\section{Conclusions}
Light-ray trajectories are known to get bent in a gravitational field, an observational confirmation of which is seen in the bending in Sun's gravitational field, for light rays coming from distant astronomical objects and passing close to Sun's limb. 
Here we determined, from a theoretical perspective, not only the exact shapes of the bent trajectories of light rays, emitted  isotropically by a source supported in a gravitational field, we also demonstrated that even the electric field lines of a charge get bent due to gravity. Using the strong principle of equivalence, we showed how the electrostatic field of a 'supported' charge gets bent in a gravitational field. It was, in fact, shown that the electric field lines follow exactly the trajectories of light rays emitted initially isotropically in a gravitational field.  
The gravitational bending of electric field lines of a 'supported' charge, like that of the light rays from a source stationary in the gravitational field, increases with depth in the gravitational field, becoming almost vertical as the horizon is approached. 
It was further shown that the field lines from a freely falling charge in a gravitational field are radial from the instantaneous position of the freely falling charge, but are bunched toward the horizontal plane, normal to the direction of free fall. 
We showed that while for a freely falling charge there is no change in the total electric flux through a plane either above and below, but on the other hand  
for a charge stationary in a gravitational field, due to the bending of electric field lines, the electric flux is less  through a horizontal plane ``above'' the charge than that in a plane ``below'' the charge, contrary to what would be expected in the absence of the gravitational field. As the bending of electric field lines becomes much more acute, for a charge held nearer  to the horizon, much more electric flux passes through a narrower region in the plane ``below'' and much lesser flux through a similar region in the plane ``above''. 
An interesting inference drawn from the gravitational bending of electric field lines is the presence of a finite electric field inside a uniformly charged spherical shell stationary in a gravitational field, a suitable experimental follow up of which might lead to a possible test of the strong principle of equivalence.
\section*{Declarations}
The author has no conflicts of interest/competing interests to declare that are relevant to the content of this article. 
No funds, grants, or other support of any kind was received from anywhere for this research.
\section*{Appendices}
\appendix
\section{Finite electric field inside a uniformly charged spherical shell held stationary in a  gravitational field}\label{SAA}
One interesting consequence of the gravitational bending of the electrostatic field lines is that 
an ``upward'' force due to bending of electrostatic field lines could make an electric dipole to float against gravity in a uniform gravitational field \cite{Bo79,Gr86}. Another consequence could be the presence of a finite electric field inside a uniformly charged spherical shell, when held in a gravitational field.

Ordinarily a sphere of radius $r_0$ having a charge $Q$, distributed uniformly over its surface and permanently stationary in an inertial frame, has a simple radial Coulomb electric field, whose value on the surface, outside the spherical shell, is
\begin{equation}
\label{eq:38a.0}
{\bf E}_{\rm c}= \frac{Q\:{\bf n}}{r_0^{2}}\:,
\end{equation} 
while the field inside the shell is zero. Here ${\bf n}$ is the outward radial unit vector at the surface of the sphere. 
 
However the same uniformly charged spherical shell, when undergoing a uniform acceleration $\bf a$, assuming the radius of the sphere to be small so that $r_{\rm o}\ll c^2/a$, has a constant, finite electric field inside the spherical shell, as well as on the shell, whose value to the lowest order in $r_{\rm o}$ is \cite{24,Er04,68,15}
\begin{equation}
\label{eq:38a.1}
{\bf E}_{\rm a}=-\frac{2Q{\bf a}}{3r_{\rm o} c^{2}}\:.
\end{equation}
It follows from the strong principle of equivalence that a sphere with a uniform surface 
charge density, but supported in a gravitational field, ${\bf g}_0$ say, on the
surface of earth, has a constant, finite electric field {\em inside} it,  as well as on the surface of the shell 
\begin{equation}
\label{eq:38a.3}
{\bf E}_{\rm g}=\frac{2Q{\bf g}_0}{3r_{\rm o} c^{2}}\:,
\end{equation}
along the direction of ${\bf g}_0$ ($=-{\bf a}$), the acceleration due to gravity. 
Then just outside the surface of a uniformly charged sphere the electric field is,
\begin{equation}
\label{eq:38a.2}
{\bf E}=\frac{{Q\bf n}  }{r_{\rm o} ^{2}}+\frac{2Q{\bf g}_0}{3r_{\rm o} c^{2}}\:.
\end{equation}
Except for the first (Coulomb field) term in Eq.~(\ref{eq:38a.2}) which is radial (along $\bf n$), the electric field given by the second term, resulting from a  bending of electrostatic field lines due to gravity, is continuous across the surface and is constant, to this order, both in direction and magnitude, at all points inside as well as on the charged sphere,  and is along ${\bf g}_0$. 
The spherical shell, due to its uniform charge density $\sigma=Q/4\pi r_0^2$, has an outward repulsive force, $2 \pi \sigma ^{2}$ per unit area, on every surface element \cite{PU85,Si92}, however due to the spherical symmetry, the consequential net self-force on the whole shell becomes zero. On the other hand, due to the second term in Eq.~(\ref{eq:38a.2}), there is a finite gravity-induced self-force 
\begin{equation}
\label{eq:38a.4}
{\bf F}_{\rm g}=\frac{2Q^2{\bf g}_0}{3r_{\rm o} c^{2}}\:,
\end{equation}
along ${\bf g}_0$, on the shell. 
\begin{figure}[t]
\begin{center}
\includegraphics[width=\columnwidth]{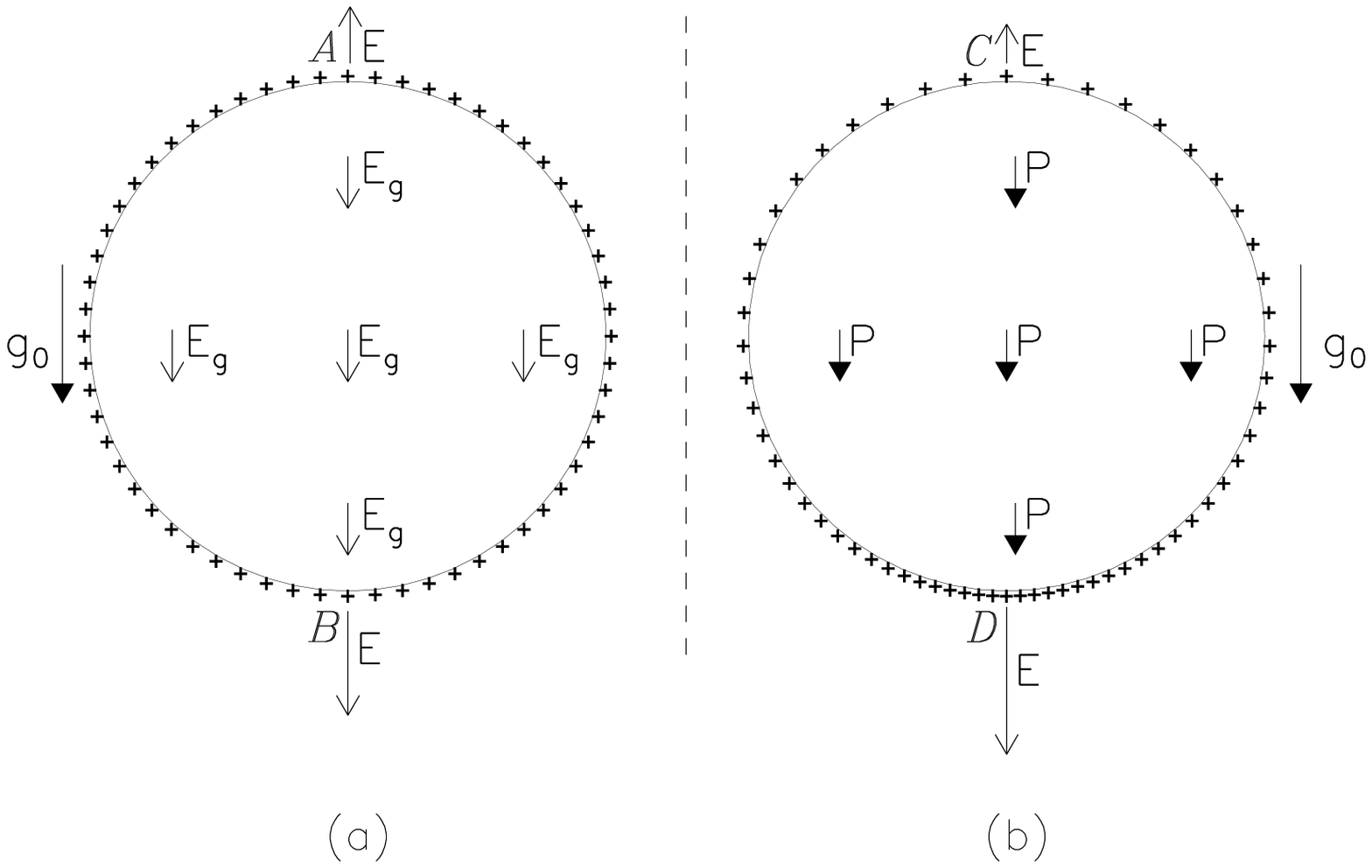}
\caption{A schematic representations of the effect of gravity on the electrostatic field. (a) A sphere, made of a non-conducting material and having a uniform surface charge density, is supported in a uniform gravitational field $g_0$. The gravity gives rise to an electric field ${E}_{\rm g}$,  along $g_0$, uniform both inside the shell and on the spherical surface, that however, adds to the electric field strength at point $B$ (``under'' the sphere) and subtracts from the electric field at point $A$ (``above'' the sphere). 
(b) A sphere made of a conducting material and having initially a uniform surface charge density, when supported in a uniform gravitational field, undergoes a redistribution of the surface charge density, equivalent to a uniform polarization density ${P}$ along ${g_0}$ within the spherical volume, whose electric field cancels ${E}_{\rm g}$ on the conducting surface as well as inside the shell. The electric field due to the surface charge density further adds to the electric field strength at point $D$ (``under'' the sphere) and subtracts from the electric field at point $C$ (``above'' the sphere).}
\label{F3}
\end{center}
\end{figure}
\subsection{Spherical shell of a non-conducting material}\label{SAAA}
Figure~\ref{F3}a is a schematic representations of the electric field of a sphere made of a non-conducting material, having a uniform distribution of charge over the surface, supported in a uniform gravitational field. 
Such gravity-induced electrostatic field could in principle be a test of the strong principle of equivalence, though a detection of such a field {\em inside} a finite-sized
sphere could be quite difficult because of the weak nature of this field ($\propto {g}_0/c^{2}$) as compared to the effects of any non-uniformity in the spherical distribution (whose effects perhaps could partially be eliminated say, by a 180 degree rotation of the sphere, without disturbing the charge distribution). For one thing the sphere will necessarily
have to be made of a highly non-conducting material to avoid
cancellation of the gravity dependent inside-electric fields 
by a redistribution of the conduction electrons. However, from Eq.~(\ref{eq:38a.2}), the electric field just ``under'' the sphere (at point $B$ in Fig.~\ref{F3}a) will be stronger than that just ``above'' (at point $A$ in Fig.~\ref{F3}a) and also could in principle be another variant of the test of the strong principle of equivalence, though the practical difficulties
could be immense, in this case too.
\subsection{Spherical shell of a conducting material}\label{SAAB}
Figure~\ref{F3}b shows the electric field of a sphere made of a conducting material, initially with a uniform distribution of charge over the surface. When supported in a uniform gravitational field,  due to gravity, a constant electric field, ${E}_{\rm g}$, both inside as well as on the spherical shell would be generated. This in turn would give rise to a redistribution of the surface charges of the conducting sphere so as to cancel ${E}_{\rm g}$ on the conducting surface as well as inside the shell. An excess charge density $\sigma =-3{E}_{\rm g}\cos\theta/4\pi$ over the sphere, $\theta$ being the polar angle with respect to the ``up'' direction along the $z$-axis, gives rise to a uniform electric field $-2Q{\bf g}_0/3r_{\rm o} c^{2}$, inside as well as on the spherical conducting surface \cite{PU85},
which would cancel ${\bf E}_{\rm g}$ due to gravity (Eq.~(\ref{eq:38a.3})), on the conducting surface as well as inside the shell.

This excess charge density is equivalent to a uniform electric polarization density ${\bf P}=3{\bf E}_{\rm g}/4\pi$ along ${\bf g}_0$, within the spherical volume, whose electric field not only cancels ${E}_{\rm g}$ on the conducting surface as well as inside the shell, 
it further gives rise to an external field whose strength at the ``top'' (point $C$) as well as at the ``bottom'' (point $D$) of the spherical surface (Fig.~\ref{F3}b) will be \cite{PU85}
 \begin{equation}
\label{eq:38a.5}
{\bf E}=\frac{8\pi{\bf P}}{3}=\frac{4Q{\bf g}_0}{3 r_{\rm o} c^{2}}\:.
\end{equation}
This will make the field strength ``under'' the sphere (at point $D$ in Fig.~\ref{F3}b) even stronger to $E_{\rm c}+4\pi P$, while that ``above'' (at point $C$ in Fig.~\ref{F3}b) will make it still weaker to $E_{\rm c}-4\pi P$. Even though a low value of $g$ on the surface of earth might appear to be a deterrent, but a suitable variant of such an experiment, with a clever design, might make it a possible test of the strong principle of equivalence.
Some other aspects of the experimental feasibilities of detecting the gravity-induced forces on a charged spherical shell, supported in a gravitational field, can be found in \cite{Ma15}.
\section{Poynting flow for a freely falling charge}\label{SAB}
From Eq.~(\ref{eq:38.03a}), the radial component of the Poynting vector for a freely falling charge is zero everywhere  
\begin{equation}
\label{eq:38a.04}
\hat{\bf R} \cdot {\cal S}= \frac{c}{4\pi}\hat{\bf R} \cdot ({\bf E}\times {\bf B})=0\;,
\end{equation}
which is consistent with no radiated power going out from the freely falling charge.

There is, however, a finite Poynting flow along $\bf v$, the direction of motion of the charge due to its free fall.
\begin{equation}
\label{eq:38a.05}
S_{\rm z}=\frac{c}{4\pi}E_{\rho}B_{\Phi}=-\frac{v}{4\pi}E^2_{\rho}
\end{equation}
\begin{figure}[t]
\begin{center}
\includegraphics[width=\columnwidth]{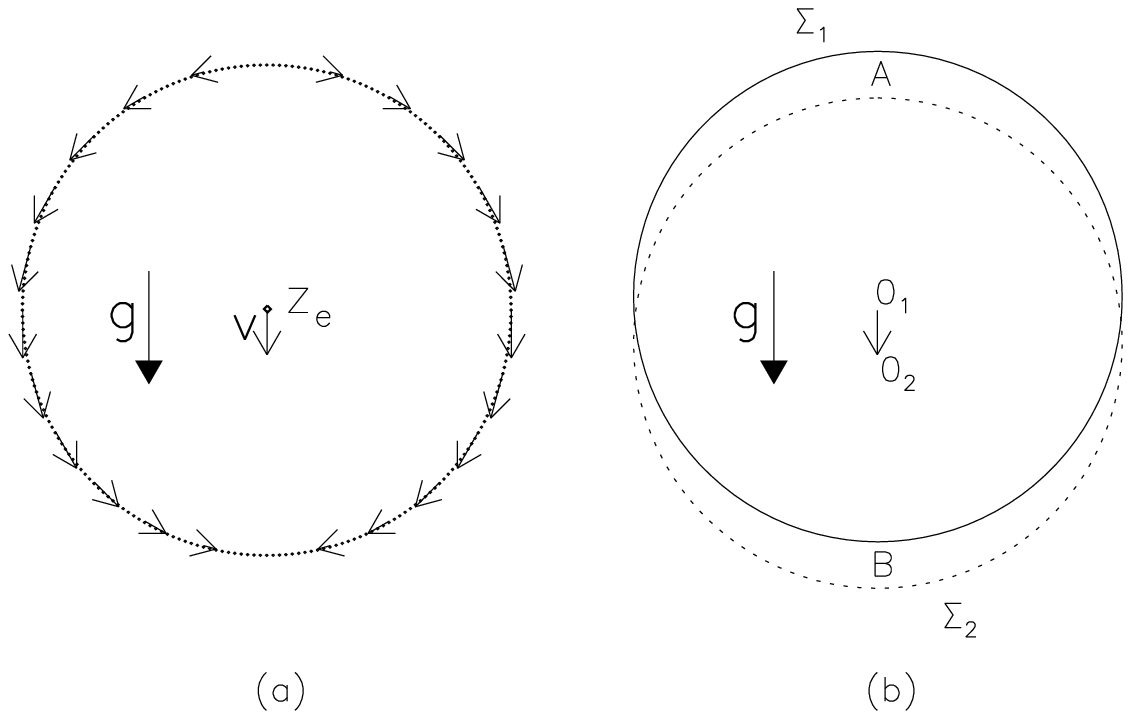}
\caption{(a) A schematic representation of the Poynting vectors around the instantaneous position $z_{\rm e}$ of the freely falling charge in a gravitation field of strength $\bf g$. The Poynting vectors on a sphere are tangential to the spherical surface with the overall Poynting flow along the velocity $v$ of the freely falling charge, representing the convective flow of the self-fields along with the moving charge. (b) As the charge falls with a velocity $v$ from its position $O_1$ to $O_2$, its self-fields also move with it. If we consider two spherical volumes, $\Sigma_1$ and $\Sigma_2$ around the two charge positions, the field energy in the region of intersection $B$ between the two spheres increases at the cost of the field energy in the region $A$, where it decreases. The Poynting vectors, seen in (a), represent the flow of the self-field energy from region $A$ to $B$ with the charge movement from $O_1$ to $O_2$.}
\label{F5}
\end{center}
\end{figure}

The overall Poynting flow here is directly proportional to the instantaneous velocity of the charge, along $-\hat{z}$ direction, and represents the convective flow of the self-fields along with the freely falling charge (Fig.~\ref{F5}a). As the charge falls with a velocity $v$ from its position $O_1$ to $O_2$, its self-fields also move with it (Fig.~\ref{F5}b). If we consider two spherical volumes, $\Sigma_1$ and $\Sigma_2$ around the two charge positions, the field energy in the region of intersection $B$ between the two spheres increases at the cost of the field energy in the region $A$, where it diminishes. The Poynting flow seen in Fig.~\ref{F5}a is due to the flow of the self-field energy from region $A$ to $B$ as the charge moves from $O_1$ to $O_2$.

Thus, from an explicit computation of Poynting vectors, we have seen that for a freely falling charge, no electromagnetic power is being radiated away from the charge. Instead, the net Poynting flow is in the direction of movement of the charge, indicating a convective flow of energy in the fields surrounding the moving charge, essentially a 'downward' transport of the self-fields along with the freely falling charge.
\section{Beaming along a forward direction is only when considered with respect to the time-retarded positions of an emitting source}\label{SAC}
\begin{figure}
\begin{center}
\includegraphics[width=\columnwidth]{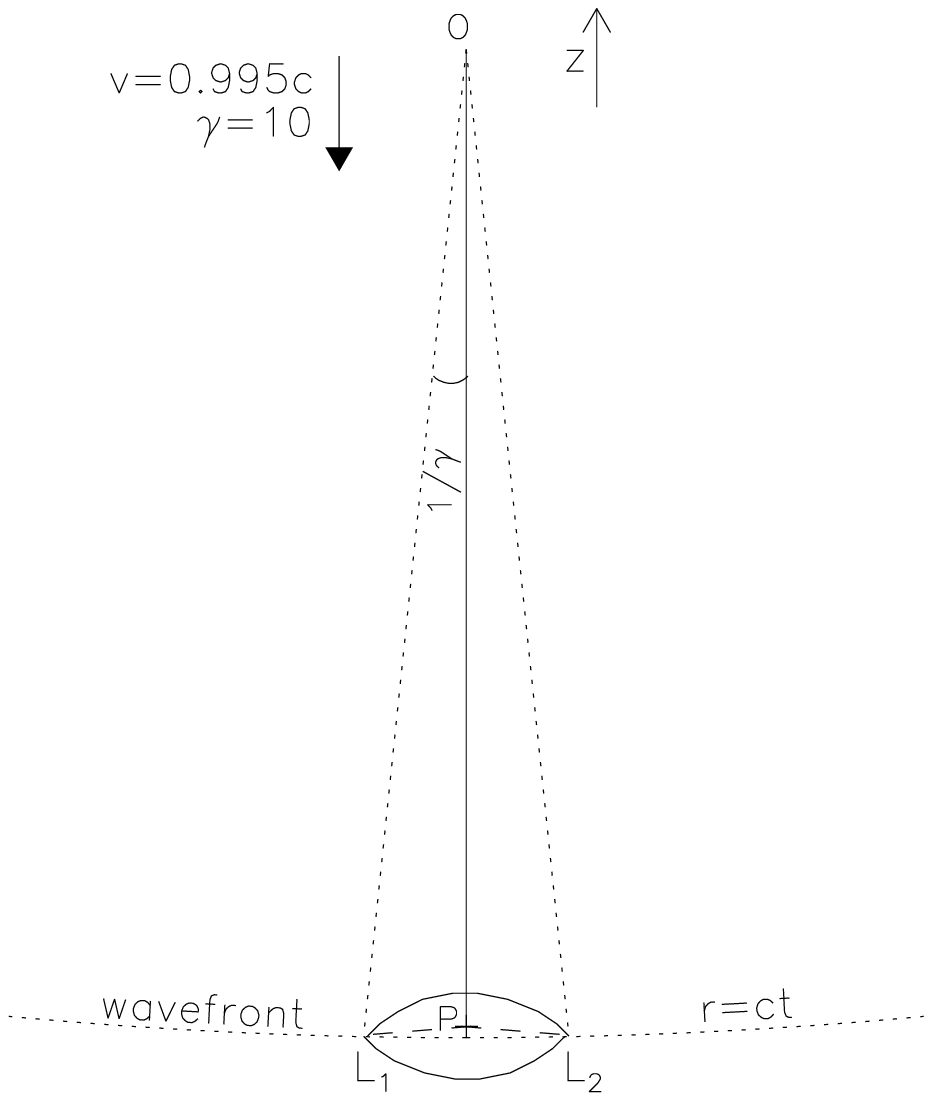}
\caption{A schematic representation of the angular distribution of the flux of light rays with respect to the time-retarded position $O$ of the source ${\cal S}$, moving 'downward' with a relativistic velocity $v=0.995c$ and the corresponding Lorentz factor $\gamma=10$. 
One half of the total number of light rays lies within a cone of half opening angle $\theta \sim 1/\gamma$, about the direction of motion of the source. 
During the time interval $t$, while a light ray from the retarded-time position $O$ moving along $\theta$, arrives at $L_1$ on the wavefront $r=ct$, the source ${\cal S}$ meanwhile moves from $O$ to $P$, a distance $vt=ct(1-1/2\gamma^2)\approx ct(1-\theta'^2/2)\approx ct \cos\theta$. Therefore $OP= OL_1\cos\theta$, implying $OPL_1$ is a right angle triangle. With respect to instantaneous position $P$ of the source ${\cal S}$, the rim of the cone of the half opening angle $\theta \sim 1/\gamma$ thus lies in a horizontal plane, perpendicular to the direction of motion.}
\label{F8}
\end{center}
\end{figure}

Let a source ${\cal S}$ of light be freely falling in the equivalent gravitational field of a comoving accelerated frame ${\cal A}$ and that the source ${\cal S}$ has velocity $v$ due to free fall at some instant of coordinate time in ${\cal A}$. Let an inertial frame, ${\cal I}$, be instantaneously coincident with the comoving accelerated frame ${\cal A}$ at that instant, from which we want to consider the position of the source and its emitted light rays in frame ${\cal A}$. The freely falling source, by the equivalence principle, can be considered to be at rest in an inertial frame, say ${\cal I'}$, and let the source ${\cal S}$ be emitting light isotropically in that frame. Then the  frame ${\cal I}'$, along with the source ${\cal S}$, is moving relative to ${\cal I}$ with the velocity $v$ along $-z$ direction.  For a relativistic velocity, $v \sim c$ or $\gamma \gg 1$, light rays in ${\cal I}'$ moving along $\theta'=90^\circ$, with $\theta'$ measured with respect to the $z'$-axis, will appear in ${\cal I}$ to be moving along $\theta\sim 1/\gamma$ about the direction of motion of ${\cal S}$ \cite{RG70}. Accordingly, the light rays with $\theta'>90^\circ$, i.e. one half of all light rays in ${\cal I}'$, will appear in frame ${\cal I}$ to be lying in a narrow cone of half opening angle $\theta \sim 1/\gamma$ around the $-z$ axis. Figure~\ref{F8} shows a schematic representation of this, where lines $OL_1$ and $OL_2$ demark the cone of half opening angle $\theta \sim 1/\gamma$, within which one half of the total light rays lie. The apex of the cone is the corresponding time-retarded position $O$ of the moving source. 

During a time interval $t$, while the light rays emitted at $O$ and moving along $OL_1$ or $OL_2$ cover a distance $r=ct$, the source ${\cal S}$ moving with a relativistic velocity  $v\approx c(1-1/2\gamma^2)$, is not lagging far behind the spherical wavefront at $r=ct$. Then the distance $OP$, moved by the source, $ct(1-1/2\gamma^2)\approx ct(1-\theta^2/2)\approx ct \cos \theta$, indicates that $OPL_1$ or $OPL_2$ are right angle triangles.
Thus the rim of the cone with respect to the instantaneous position $P$ of the source ${\cal S}$, seems to be in a plane perpendicular to $OP$, the direction of motion.

Actually, even for a non-relativistic motion, a light ray  along $\theta'=90^\circ$ in ${\cal I}'$, will appear in ${\cal I}$, to be along $\theta$, given by the aberration formula (Eq.~(\ref{eq:38.ab6}) with $\theta_0=\theta'$), $\cos\theta=-v/c$ or $\sin\theta=1/\gamma$. The source has thus moved a distance, $vt=ct\cos\theta$, and the cone rim accordingly, represented by points  $L_1$ or $L_2$, and enclosing half of the total number of light rays, lies in the horizontal plane passing through the `present' position $P$ of the source on the $z$-axis (Fig.~\ref{F8}). With increasing $t$, as the wavefront moves forward with $r=ct$, the source too moves a distance $vt=ct\cos\theta$, with the rim of the cone and point $P$ always in the same horizontal plane, as seen in Fig.~\ref{F7}b. Thus in frame ${\cal I}$, light rays going along $\sin\theta=1/\gamma$ with respect to the retarded-time position $O$ of the source will be moving in a horizontal plane (perpendicular to the $z$-axis) passing through the instantaneous position $P$ of the source. It should be noted that here we are considering the situation in frame ${\cal I}$, in which the source is moving with a velocity $v$ and the  flux of light rays, emitted from the time-retarded position $O$ of the source, is distributed about the horizontal plane passing through $P$,  the instantaneous 'present' position of the source. 

However, it still remains to be demonstrated that the light rays would be increasingly bunched around the horizontal plane passing through the instantaneous present location of the source, as the velocity of the freely falling source approaches $c$, the velocity of light (Fig.~\ref{F7}b). In order to compute the light-ray trajectories with respect to the present location of the source, one can consider a light ray emitted along a rod extending radially at a fixed orientation, say along $\theta'$ with respect to the $z'$-axis, in ${\cal I}'$. Different light rays will be along different angles, like the spokes of a wheel. Then the light-ray trajectories with respect to the source position in frame ${\cal I}$ too would be along the corresponding orientations of the rods in that frame. However, the rods, being stationary in frame ${\cal I}'$, are in motion in frame ${\cal I}$ and the moving rod lengths along the direction of motion, i.e., their $z$-components, should undergo Lorentz contractions by a factor $\gamma$. Accordingly, the orientations of rods in frame ${\cal I}$, as well as in ${\cal A}$  that has ${\cal I}$ as its instantaneously coincident inertial frame, will be related by $\tan \theta=\gamma \tan \theta'$ to those in frame  ${\cal I}'$. Therefore, with higher and higher $\gamma$ for the increasing velocity of the freely falling source ${\cal S}$, the rods as well as the light ray trajectories in frame ${\cal A}$ will be increasingly tipped toward the horizontal plane, which is at $90^\circ$ to the $z$-axis.

More formally, Eq.~(\ref{eq:38.ab5}) gives the Lorentz transformation from ${\cal I}'$ to ${\cal I}$ for the light ray movement between two events, say ${\cal E}_1$ and  ${\cal E}_2$. But the difference now is that the source ${\cal S}$, instead of being stationary in the gravitational field, is freely falling. However, it is stationary in the co-falling inertial frame ${\cal I}'$ and with respect to ${\cal I}$, which is instantaneously coincident with  ${\cal A}$, moves a distance $OP=v\Delta t$ along $-z$ direction, during the time interval $\Delta t$. Then the light ray, as seen in ${\cal I}$ with respect to the `present' position $P$ of the source ${\cal S}$, is located at   
\begin{eqnarray}
\label{eq:38.ab8}
\nonumber
{\Delta z_1}&=& \Delta z+ v\Delta t\\
\nonumber
&=& {\gamma(\Delta z'- v\Delta t'})+v \gamma(\Delta t' - v \Delta z'/c^2)\\
&=&\Delta z' /\gamma
\,,
\end{eqnarray}
which yields
\begin{eqnarray}
\label{eq:38.ab9}
{\tan\theta}= {\Delta \rho}/\Delta z_1= \gamma{\Delta \rho'}/\Delta z'= \gamma\tan \theta'\,.
\end{eqnarray}
Thus a light ray emitted along angle $\theta'$ in frame ${\cal I}'$ will appear to be moving along  angle $\theta$ (given by Eq.~(\ref{eq:38.ab9})) with respect to the `present' position of the freely falling source in ${\cal I}$ as well as in ${\cal A}$. 

To summarize, light rays from a freely falling source, emitted in the horizontal plane (perpendicular to the direction of motion) in the rest frame ${\cal I}'$ of the source, will be moving in the horizontal plane in ${\cal I}$ also. However, the emitted light rays having an isotropic distribution about the source position in frame ${\cal I}'$, when considered in frame ${\cal I}$ with respect to the {\em time-retarded position} of the emitting source, will be beamed within a narrow angle $\sim 1/\gamma$ around the forward direction of motion of the source, nonetheless would be bunched in a narrow angle $\sim 1/\gamma$ about the horizontal plane through the {\em present position} of the source. Then carrying over these results to the comoving accelerated frame ${\cal A}$, having ${\cal I}$ as its instantaneously coincident inertial frame, and further extending them to observers in the gravitational field, using the equivalence principle, light rays trajectories from a freely falling source in a gravitational field, radial from the instantaneous position of the freely falling source, are bunched toward the horizontal plane, normal to the direction of free fall, as seen in Fig.~\ref{F7}b.


\end{document}